# Graphene-Based Hole Selective Layers for High-Efficiency, Solution-Processed, Large-Area, Flexible, Hydrogen-Evolving Organic Photocathodes

*S. Bellani*[a]‡, *L. Najafi*[a]‡*, B. Martín-García*[b]*, A. Ansaldo*[a]*, Antonio E. Del Rio Castillo*[a]*, M. Prato*[c]*, I. Moreels*[b] *and F. Bonaccorso*[1]*

[a] Graphene Labs, Istituto Italiano di Tecnologia, via Morego 30, 16163 Genova, Italy.
[b] Nanochemistry, Istituto Italiano di Tecnologia, via Morego 30, 16163 Genova, Italy.
[c] Materials Characterization Facility, Istituto Italiano di Tecnologia, via Morego 30, 16163 Genova, Italy.

‡ S. Bellani and L. Najafi contributed equally. All authors have given approval to the final version of the manuscript.

* Corresponding author: francesco.bonaccorso@iit.it.

## ABSTRACT

Regio-regular poly(3-hexylthiophene-2,5-diyl) (rr-P3HT), the work-horse of organic photovoltaics, has been recently exploited in bulk heterojunction (BHJ) configuration with phenyl-C61-butyric acid methyl ester (PCBM) for solution-processed hydrogen-evolving photocathodes, reaching cathodic photocurrents at 0 V *vs.* RHE ($J_{0V\ vs\ RHE}$) of up to 8 mA cm$^{-2}$. The photoelectrochemical performance of these photocathodes strongly depends on the presence of the electron (ESL) and hole (HSL) selective layer. While TiO$_2$ and its sub-stoichiometric phases are consolidated ESL materials, the currently used HSLs (*e.g.,* MoO$_3$, CuI, PEDOTT:PSS, WO$_3$) suffer electrochemical degradation under hydrogen evolution reaction (HER)-working conditions. In this work, we use solution-processed graphene derivatives as HSL to boost efficiency and durability of rr-P3HT:PCBM-based photocathodes, demonstrating record-high performance. In fact, our devices show cathodic $J_{0V\ vs\ RHE}$ of 6.01 mA cm$^{-2}$, onset potential ($V_o$) of 0.6 V *vs.* RHE, ratiometric power-saved efficiency ($\varphi_{saved}$) of 1.11% and operational activity of 20 hours in 0.5 M H$_2$SO$_4$ solution. Moreover, the designed photocathodes are effectively working in different pH environments ranging from acid to basic. This is pivotal for their exploitation in tandem configurations, where photoanodes operate only in restricted electrochemical conditions. Furthermore, we demonstrate the scalability of our all-solution processed approach by fabricating a large-area (~9 cm$^2$) photocathode on flexible substrate, achieving remarkable cathodic $J_{0V\ vs\ RHE}$ of 2.8 mA cm$^{-2}$, $V_o$ of 0.45 V *vs.* RHE and $\varphi_{saved}$ of 0.31%. This is the first demonstration of high-efficient rr-P3HT:PCBM flexible photocathodes based on low-cost and solution-processed manufacturing techniques.

## INTRODUCTION

Strong economic growth and expanding populations are currently driving the increase of the world energy consumption.[1,2] The U.S. Energy Information Administration's recently released International Energy Outlook 2016 (IEO2016) projects outlining a world energy consumption growth by 48% between 2012 and 2040.[3] In this framework, oil, natural gas and coal will continue to meet about 80% of the global energy demand, rising climate disruption concerns fuelled by $CO_2$ emission.[4-8] Sunlight-powered hydrogen production through artificial photosynthesis represents a promising method for tackling the fuel demand in a post-fossil era.[8-10] In fact, hydrogen, while keeping many of the advantages of hydrocarbon fuels, avoids the drawback of $CO_2$ emissions upon combustion.[6-9] This paves the way for the development of the so-called *Hydrogen Economy*,[11,12] which refers to the vision of using only hydrogen for both energy conversion[13] and storage,[14] being already a reality in the municipality of Utsira, Norway.[15]

In the implementation of hydrogen production on a global scale, different methods have been investigated,[6-9] such as hydrocarbon routes[16,17] (*e.g.*, natural gas steam reforming,[16-18] coal gasification[16,17,19]) and water-splitting processes[9,16,17,20] (*e.g.*, photovoltaics -PV-[21,22] and wind-electrolysis,[9,23] photolysis,[24-26] thermolysis[24-27] and photoelectrochemical (PEC) water-splitting[6-9,28]). Amongst them, direct PEC conversion of sunlight into hydrogen and oxygen by water-splitting represents the most scalable and cost-effective solution.[29-32] A water-splitting PEC cell comprises a semiconductor photoelectrode and a counter electrode immersed in an aqueous electrolyte.[33,34] Semiconductor photoelectrodes absorb light photogenerating electrical charges,[33,34] which are needed to perform the redox chemistry of the hydrogen evolution reaction (HER: $4H^+ + 4e^- \rightarrow 2H_2$) and oxygen evolution reaction (OER: $2H_2O \rightarrow O_2 + 4H^+ + 4e^-$).[35,36] The electrochemical potential of the bottom of the photoelectrode conduction band (CB) must be more negative than the $H^+/H_2$ redox level ($E^0_{H^+/H_2}$ = 0 V), while the one of the top of the photoelectrode valence band (VB) must be more positive than the $O_2/H_2O$ redox level ($E^0_{O_2/H_2O}$ = 1.23 V).[29,37] These thermodynamic constraints limit the choice of the semiconductor materials to the ones having band gap exceeding 1.23 V.[30-32,38] Thus, these single semiconductor absorbers cannot harvest a significant portion of the solar spectrum and therefore their potential solar-to-hydrogen conversion efficiency ($\eta_{STH}$) is intrinsically limited (Shockley-Quiesser limit[39]).[40-42] Nevertheless, tandem PEC cells based on two vertically stacked absorbing materials with different band gap can simultaneously optimize the solar light harvesting[41] and increase the photovoltage,[42-46] which in turn enhances the photocurrent values.[43-46] Currently, tandem cells with $\eta_{STH}$ up to 18% have been demonstrated,[47] mainly using compound III-V semiconductors.[44-46,48-51] However, the manufacturing cost of these materials is significantly higher (*e.g.*, their PV-module cost are > 2 USD/W$_p$)[52] than *e.g.*, Si (PV-module cost between 0.5-1 USD/W$_p$).[53] Recently, $\eta_{STH}$ >10% has been demonstrated by using cheaper materials[47] such as Si,[54] CIGS[55] and halide perovskites.[56,57]

Despite these results, the main obstacles for the commercialization of water-splitting PEC devices include the use of expensive and not scalable deposition techniques (such as atomic layer deposition,[58-62] ion layer adsorption and reaction,[63-65] sputtering[66-69] and evaporation of metal/metal oxide protective layers),[70,71] and limited lifetime of devices in contact with aqueous electrolytes.[47,50,51] The latter, in particular, is challenging for the implementation of monolithically integrated devices fully immersed in water.[48,49,69,72]

Organic conjugated polymers have been proposed as candidate photocathode materials[73-85] due to their low costs[86] (e.g., their potential PV module costs are ~1 USD/$W_p$[86-88]) and compatibility with high-throughput production techniques (solution-processed roll-to-roll and large-area deposition processes).[90-92] In particular, regio-regular poly(3-hexylthiophene-2,5-diyl) (rr-P3HT) has been recently exploited in bulk heterojunction (BHJ) configuration with phenyl-C61-butyric acid methyl ester (PCBM) for photocathodes reaching cathodic photocurrents at 0 V vs. reversible hydrogen electrode (RHE) scale ($J_{0V\ vs\ RHE}$) of 8 mA cm$^{-2}$ and onset potential ($V_o$) (defined as the potential at which the photocurrent related to the HER is observed) of 0.7 V vs. RHE.[81,83] The rr-P3HT has a direct bandgap of 1.9 eV,[93,94] thus close to the optimum value for a PEC tandem device ($\eta_{STH}$ of 21.6% is predicted stacking 1.89 eV and 1.34 eV energy band gap semiconductors).[44-46] Moreover, the lowest unoccupied molecular orbital (LUMO) level of PCBM (LUMO$_{PCBM}$) is several hundreds of millivolts more negative than the $E^0_{H^+/H_2}$ potential (LUMO$_{PCBM}$ - $E^0_{H^+/H_2}$ > -0.5 V),[95,96] thus photogenerated electrons possess the energy enabling the HER process.[97] Furthermore, the optoelectronic properties of rr-P3HT, such as light absorption and charge photo-generation, are fully retained in aqueous environments.[98-103] The architecture of rr-P3HT:PCBM-based photocathodes consists of rr-P3HT:PCBM BHJ sandwiched between two charge-selective layers (CSLs), and a thin electrocatalyst (EC) layer.[75-79,84,85] The photocathode figures of merit (FoM) strongly depends on the presence of the HSL and ESL materials.[104] While TiO$_2$ and its sub-stoichiometric phases have demonstrated to be consolidated ESL materials,[80-83,85] the choice for the HSL counterpart has been a more complex task. In fact, although efficient HSL materials have been identified (e.g., MoO$_3$,[77] WO$_3$,[82] NiO,[80] CuI,[83,85] PEDOT:PSS[75-81]) their intrinsic electrochemical degradation under HER-working conditions limited the lifetime of the photocathodes, lasting from several minutes to about few hours (up to 10 hours in the case of WO$_3$).[75-85] Moreover, the operational activity of the most efficient structures has been demonstrated only in acidic conditions,[76-85] with only a few examples showing remarkable cathodic $J_{0V\ vs\ RHE}$ of 1.2 mA cm$^{-2}$ at neutral pH.[75] The possibility to design a photocathode able to operate in a larger pH window is beneficial for the development of tandem architectures operating at neutral or alkaline solutions.[105] In these conditions, the photoanodes (having complementary electrochemical properties) of the tandem architecture usually exhibit lower overpotential loss for OER.[106,107] In addition, the possibility to operate at near-neutral pH aqueous conditions is of utmost interest, i.e., permitting the use of sea and river water as easy-available and non-hazardous/corrosive electrolyte.[108] This relaxes the stability constraints of practical photoactive and catalyst components.[109,110]

The research of novel HSL materials for rr-P3HT:PCBM-based photocathodes has recently involved two dimensional (2D) materials, including graphene derivatives, *e.g.*, graphene oxide (GO) and reduced graphene oxide (RGO),[80] and transition metal dichalcogenides (TMDs), *e.g.*, $MoS_2$.[84] For example, electrochemically p-doped $MoS_2$ flakes, produced by Li-aided exfoliation, enabled to reach cathodic $J_{0V}$ vs. RHE of 1.21 mA cm$^{-2}$ and $V_o$ of 0.56 V *vs.* RHE,[84] *i.e.,* approaching the state-of-the-art efficiency of solution-processed organic $MoO_3$-based photocathodes.[80] The advantage of using the aforementioned 2D materials is linked with the possibility of creating and designing layered artificial structures with on-demand electrochemical properties[111-114] by means of large-scale, cost-effective solution processed production methods.[115-124] In fact, the possibility to produce 2D materials from the exfoliation of their bulk counterpart in suitable liquids[125-131] permits to formulate functional inks.[132-134] The latter can then be deposited on different substrates by established printing/coating techniques[118-147] So far, the durability of the graphene/TMDs-based photocathodes has been tested over no more than 1 hour-period[80,84] and further investigations on these classes of 2D materials as CSLs for PEC application are needed. In fact, despite both graphene derivatives and TMD flakes have been demonstrated as efficient HSL and ESL in solution-processed PVs, including organic,[148-154] dye sensitized,[155,156] and perovskite solar cells,[157-161] no endurance tests have been reported for graphene/TDMs-based photocathodes.

In this work, we report on solution-processed graphene derivatives, *i.e.*, GO and RGO, as HSLs for high-efficiency solution-processed rr-P3HT:PCBM-based photocathodes with improved stability under HER-working conditions. These results are obtained by adopting two different strategies. The first one relies on the fabrication of hydrogen-bonded fluorine-doped tin oxide (FTO)/graphene-based HSL/rr-P3HT:PCBM structures through the chemical functionalization of GO/RGO (compounds here named as f-GO and f-RGO, respectively), with (3-mercaptopropyl)trimethoxysilane (MPTMS) in an ethanol solution.[162] The second one is the implementation of solution-processed conductive and catalytic Pt on carbon-tetrafluoroethylene-perfluoro-3,6-dioxa-4-methyl-7-octenesulfonic acid copolymer blend (Pt/C-Nafion) overlay. The optimization of the proposed architectures allowed us to achieve a record-high efficiency for solution-processed rr-P3HT:PCBM-based photocathodes, extending their operational activity up to 20 h. This result outperforms the state-of-the-art endurance for rr-P3HT:PCBM-based photocathodes (10 hours for photocathodes based on 100nm-thick $WO_3$ film as HSL).[82] Moreover, our photocathodes are extremely versatile, showing high PEC activity in different pH conditions, *i.e.*, ranging from acid to basic. This is pivotal for the exploitation of the proposed photocathodes in tandem configurations, whose photoanode activity is usually facilitated in alkaline condition.[105-107] Finally, we demonstrate the scalability of our approach, reporting the fabrication of a large area (9 cm$^2$) photocathode on flexible indium tin oxide (ITO)-coated poly(ethylene terephthalate) (ITO-PET) substrate.

## EXPERIMENTAL SECTION

### Synthesis of GO and RGO

Graphene oxide is synthesized from graphite flakes (Sigma Aldrich, +100 mesh ≥75% min) using a modified Hummer's method. Briefly, 1 g of graphite and 0.5 g of NaNO$_3$ (Sigma Aldrich, reagent grade) are mixed, followed by the dropwise addition of 25 mL of H$_2$SO$_4$ (Sigma Aldrich). After 4 h, 3 g of KMnO$_4$ (Alpha Aesar, ACS 99%) is added slowly to the above solution, keeping the temperature at 4 °C with the aid of an ice bath. The mixture is let to react at room temperature overnight and the resulting solution is diluted by adding 2 L of distilled water under vigorous stirring. Finally, the sample is filtered and rinsed with H$_2$O.

The RGO is produced by thermal annealing in a quartz tube (120 cm length and 25 mm inner diameter) passing through a three zones split furnace (PSC 12/--/600H, Lenton, UK). Gas flows are controlled upstream by an array of mass flow controllers (1479A, mks, USA). Under a 100sccm flow of Ar/H$_2$ (90/10 %), 100 mg of GO are heated to 100 °C for 20 min to remove the presence of water residuals. Subsequently, a ramp of 20 °C/min is used to reach 1000 °C, and stabilized at this temperature for 2 h. Finally, the oven is left to cool to room temperature.

### Functionalization of GO and RGO

The silane functionalization of GO and RGO is carried out following the procedure previously reported[162] and based on the covalent linking of silane groups to the GO and RGO oxygen functionalities. Briefly, 0.5 mg mL$^{-1}$ GO and RGO dispersions in ethanol (absolute alcohol, ≥99.8%, without additive, Sigma Aldrich) are sonicated for 30 min and subsequently functionalized by adding 250 µL of MPTMS (95%) (Sigma Aldrich) per mg of GO and RGO, refluxing at 60 °C for 15 h. The final product is obtained by subsequent washing with ethanol to remove the unreacted silane and precipitating the material by centrifugation. The functionalized GO and RGO are re-dispersed in ethanol by sonication at different concentrations (0.5, 1 and 1.5 mg mL$^{-1}$) to prepare the films.

### Fabrication of photocathodes

Photocathodes are fabricated according to the architecture FTO/HSL/rr-P3HT:PCBM/TiO$_2$/MoS$_3$, where GO, RGO, f-GO and f-RGO films are used as HSL. Architectures without HSL are also fabricated. FTO is deposited on soda-lime glass substrates (area 1×1.5 cm$^2$, sheet resistance 15 Ω/□, Dyesol). The surfaces of FTO are cleaned according to the following protocols: sequential sonication baths in deionized (DI) water, acetone, isopropanol (IPA) each lasting for 10 min and plasma cleaning in an inductively coupled reactor for 20 minutes (100 W RF power, excitation frequency 13.56 MHz, 40 Pa of $O_2$ gas process pressure, background gas pressure 0.2 Pa). For large-area flexible devices, ITO-PET substrates (sheet resistance 30 $\Omega$/□, Sigma Aldrich) are used. The cleaning protocols of surface of ITO are the same of that of FTO without the second sonication bath in acetone in order to avoid PET substrate dissolution.

Graphene oxide, RGO, f-GO and f-RGO are dispersed in ethanol by sonication at different concentration (0.5, 1 and 1.5 mg mL$^{-1}$) and deposited onto the previously treated FTO by spin coating ((WS-650Mz-23NPPB Laurell Tech. Corp. Spin coater) using a single step spinning protocol with rotation speed of 2000 rpm for 60 s. Post thermal annealing in Ar atmosphere at 150 °C for 10 min. is performed for the GO and RGO films. The organic polymer film used in all the architectures consisted in a blend of rr-P3HT, as the donor component, and PCBM, as the acceptor component (rr-P3HT:PCBM). rr-P3HT (electronic grade, $M_n$ 15000-45000, 99.995% trace metals basis, Sigma Aldrich) and PCBM (>99.5%, Nano C) are separately dissolved in chlorobenzene (ACS grade, 99.8%, Sigma Aldrich), at a weight ratio 1:1 and 25 mg mL$^{-1}$ on a polymer basis. Polymer blend solution is stirred at 40 °C for 24 h before use. Blend thin films are obtained by spin coating the rr-P3HT:PCBM solution using the following set of parameters: two step spinning protocol with rotation speeds of 800 rpm for 3 s followed by 1600 rpm for 60 s, respectively. This spin coating protocol produced a rr-P3HT:PCBM blend layer of 200 $\pm$20 nm thick, as measured with a Dektak XT profilometer (Bruker) equipped with a diamond-tipped stylus (2 mm) selecting a vertical scan range of 25 mm with 8 nm resolution and a stylus force of 1 mN, on an area of 0.25 cm$^2$.

TiO$_2$ paste (Ti-Nanoxide T-L/SC formulation, anatase particle size 15-20 nm, 3% wt, Solaronix) is deposited on top of rr:P3HT:PCBM by spin casting. Before its deposition rr:P3HT:PCBM films are treated by oxygen plasma for 30 s (20 W RF power, excitation frequency 13.56 MHz, 40 Pa of $O_2$ gas process pressure, background gas pressure 0.2 Pa) in order to increase their wettability by the TiO$_2$ dispersion. A three step spinning protocol with rotational speeds of 200 rpm for 3 s, 1000 rpm for 60 s and 5000 rpm for 30 s is used. Subsequently, the samples are dried for 12 h in air at room temperature. Post thermal annealing in a $N_2$ atmosphere is performed at 130 °C for 10 min for all the devices before catalyst deposition. The devices are completed by the deposition of a layer of Pt nanoparticles (>99.97% trace metals basis) (Sigma Aldrich) or Pt/C (20% Pt on Vulcan XC72, Sigma Aldrich) blended with Nafion (Nafion® 117 solution, 5% in a mixture of lower aliphatic alcohols and water, Sigma Aldrich) (Pt/C-Nafion) as catalyst for HER. The Pt catalyst layer is obtained by spin coating 1 mg mL$^{-1}$ Pt nanoparticles dispersion in DI water on top of the TiO$_2$. The Pt/C-Nafion layer is deposited by spin coating 5 mg mL$^{-1}$ Pt/C dispersion in DI water with the addition of 80 μL of Nafion dispersion. The dispersions are stirred overnight at room temperature and sonicated for 10 minutes before their use. Spinning protocols are identical to the one adopted for the TiO$_2$ deposition. No differences of protocols are applied for the deposition of the different layers in the case of large-area (9 cm$^2$) flexible photocathodes fabricated on ITO-PET.

## Material and devices characterization

UV-Vis absorption spectra of the GO and RGO dispersions in ethanol are collected using a Cary Varian 5000 UV-Vis spectrometer.

Raman measurements are carried out with a Renishaw 1000 using a 50x objective, a laser with an excitation wavelength of 532 nm and an incident power on the samples of 1 mW. The different peaks are fitted with Lorentzian functions. For each sample, 30 spectra are collected. 0.01 mg mL$^{-1}$ GO and RGO dispersions in ethanol are drop-casted on Si/SiO$_2$ (300 nm SiO$_2$) substrates and dried under vacuum. Statistical analysis of the relevant features is carried out by means of Origin 8.1 software (OriginLab).

Transmission electron microscopy (TEM) images are taken by a JEM 1011 (JEOL) transmission electron microscope, operating at 100 kV. 0.01 mg mL$^{-1}$ GO, RGO, f-GO and f-RGO dispersions in ethanol are drop-casted onto carbon coated Cu TEM grids (300 mesh), rinsed with deionized (DI) water and subsequently dried under vacuum overnight. Lateral dimensions of the flakes are measured using ImageJ software (NIH). Statistical TEM analysis is carried out by means of Origin 8.1 software (OriginLab).

Atomic force microscopy (AFM) images are obtained using AFM instrument MFP-3D (Asylum Research), with NSG30/Au (NT-MDT) probes in AC mode in air. Nominal resonance frequency and spring constant of NSG30/Au (NT-MDT) probes are 240-440 kHz and 22-100 N/m, respectively. The tip is a pyramid with 14-16 μm length, ~20 nm apex diameter. The images are processed with the Asylum AFM software (Version-13), based on IgorPro 6.22 (Wavemetrics). The GO, RGO, f-GO and f-RGO flakes are deposited on mica sheets (EMS) (V-1 quality) by drop-casting from a 0.1 mg mL$^{-1}$ of the corresponding dispersions in ethanol. Statistical analysis of the height profile signals, *i.e.,* the thickness of the flakes, is carried out by means of Origin 8.1 software (OriginLab).

X-ray photoelectron spectroscopy (XPS) is carried out with a Kratos Axis Ultra DLD spectrometer, using a monochromatic Al K$_\alpha$ source (15 kV, 20 mA). High-resolution scans are performed at a constant pass energy of 10 eV and steps of 0.1 eV. The photo-electrons are detected at a take-off angle $\phi = 0^o$ with respect to the surface normal. The pressure in the analysis chamber is kept below $7 \times 10^{-9}$ Torr for data acquisition. The binding energy scale is internally referenced to the Au 4f$^{7/2}$ peak at 84 eV. The spectra are analyzed using the CasaXPS software (version 2.3.16). The samples are prepared by drop-casting the 1 mg mL$^{-1}$ GO, RGO, f-GO and f-RGO dispersions onto 50 nm-Au sputtered coated silicon wafers.

Ultraviolet photoelectron spectroscopy (UPS) analysis is performed to estimate the position of the Fermi level (E$_F$) of the materials under investigation with the same equipment using a He I (21.22 eV) discharge lamp. The $E_F$ is measured from the threshold energy for the emission of secondary electrons during He I excitation. A -9.0 V bias is applied to the sample in order to precisely determine the low kinetic energy cut-off. The samples are prepared by drop-casting the materials onto 50 nm-Au sputtered coated silicon wafers.

The $E_F$ are also obtained, complementing the UPS measurements, by using a Kelvin probe (KP) system (KPSP020, KP Technologies Inc.). Samples are prepared by spin coating 1 mg mL$^{-1}$ GO, RGO, f-GO and f-RGO dispersions on FTO. The measurements are carried out in air and at room temperature. Both a clean Au surface ($E_F$ = -4.8 eV) and a graphite sample (HOPG, highly ordered pyrolytic graphite, $E_F$ = -4.6 eV) are used as independent references for the probe potential.

Scanning electron microscope (SEM) analysis is performed with a field-emission scanning electron microscope FE-SEM (Jeol JSM-7500 FA). The acceleration voltage is set at 5kV. Images are collected using secondary electron sensor for lower secondary electron (LEI) images and the in-lens sensor for upper secondary electron in-lens (SEI) images. Energy-dispersive X-ray (EDX) spectroscopy images are acquired at 5kV by a silicon drift detector (Oxford Instruments X-max 80) having an 80mm$^2$ window. The EDX analysis is performed using Oxford Instrument AZtec 3.1 software.

## Photoelectrochemical characterization

Photoelectrochemical measurements are carried out at room temperature in a flat-bottom fused silica cell under a three-electrode configuration using CompactStat potentiostat/galvanostat station (Ivium), controlled via Ivium's own IviumSoft. A Pt wire is used as the counter-electrode and saturated KCl Ag/AgCl is used as the reference electrode. Measurements are performed in 50 mL of different aqueous solutions at different pH values. The pH = 1 solution consists of 0.5 M $H_2SO_4$ (99.999% purity, Sigma Aldrich). The pH = 4 solution consists of sodium acetate/acetic acid (ACS reagent, >99.7%, Sigma Aldrich) buffer. The pH = 7 solution is potassium dihydrogen phosphate/di-sodium hydrogen phosphate buffer (Fluka). The pH = 10 solution consists of di-sodium tetraborate/sodium hydroxide buffer (Sigma Aldrich). Oxygen is purged from electrolyte solutions by flowing $N_2$ gas throughout the liquid volume using a porous frit for 30 minutes before starting measurements. A constant $N_2$ flow is maintained afterwards for the whole duration of experiments, to avoid re-dissolution of molecular oxygen in the electrolyte. Potential differences between the working electrode and the Ag/AgCl reference electrode are converted to the RHE scale via the Nernst equation (Equation (1)):

$$E_{RHE} = E_{Ag/AgCl} + 0.059pH + E^0_{Ag/AgCl} \quad (1)$$

where $E_{RHE}$ is the converted potential versus RHE, $E_{Ag/AgCl}$ is the experimental potential measured against the Ag/AgCl reference electrode, and $E^0_{Ag/AgCl}$ is the standard potential of Ag/AgCl at 25 °C (0.1976 V).

A 300 W Xenon light source LS0306 (Lot Quantum Design), equipped with AM1.5G filters, is used to simulate 1.5AM solar illumination (1 Sun, 100 mW cm$^{-2}$) at the surface of the samples inside the test cell (illumination area of 1 cm$^2$). Linear Sweep Voltammetry (LSV) is used to evaluate the response of devices in the dark and under 1.5AM illumination condition. The voltage is swept starting from potential more positive than $V_o$ of the photocathodes to a negative potential of -0.2 V *vs.* RHE at a scan rate of 10 mV s$^{-1}$. The main Figures of Merit (FoM) extracted from the voltammograms are: the $J_{0V\ vs\ RHE}$, the $V_o$, the maximum power point ($V_{mpp}$) (defined as d(JV)/dV=0) $V_{mpp}$, the fill factor (FF) (defined as the ratio of maximum obtainable power to the product of the $J_{0V\ vs\ RHE}$ and $V_o$ ($J_{mpp}$x$V_{mmp}$/$J_{0V\ vs\ RHE}$x$V_o$, where $J_{mpp}$ is the current density at V =$V_{mpp}$), the ratiometric power-saved relative to a non-photoactive (NPA) dark electrode with an identical catalyst (C) ($\Phi_{saved,NPA,C}$) and the ratiometric power-saved relative to an ideally non-polarizable reversible hydrogen electrode, *i.e.,* the RHE, ($\Phi_{saved,ideal}$). $\Phi_{saved,NPA,C}$ is calculated by Equation (2):

$$\Phi_{saved,NPA,C} = \eta_F \times |j_{photo,m}| \times [E_{light}(J_{photo,m}) - E_{dark}(J_{photo,m})] / P_{in} = \eta_F \times |j_{photo,m}| \times V_{photo,m} / P_{in} \qquad (2)$$

where $\eta_F$ is the current-to-hydrogen faradaic efficiency assumed to be 100 %, $P_{in}$ is the power of the incident illumination and $j_{photo,m}$ and $V_{photo,m}$ are the photocurrent and photovoltage at the $V_{mpp}$, respectively. $j_{photo}$ is obtained by calculating the difference between the current under illumination of a photocathode and the current of the corresponding catalyst. The photovoltage $V_{photo}$ is the difference between the potential applied to the photocathode under illumination ($E_{light}$) and the potential applied to the catalyst electrode ($E_{dark}$) to obtain the same current density. The subscript "m" stands for "maximum". $\Phi_{saved,NPA,C}$ reflects the photovoltage and photocurrent of a photocathode independently from the over-potential requirement of the catalyst. It is assumed that the catalyst film deposited on FTO is identical to the one deposited on TiO$_2$. $\Phi_{saved,ideal}$ is simply obtained by considering RHE as catalyst electrode, *i.e.,* setting $E_{dark}$ = 0 V *vs.* RHE in Equation (2).

Stability tests are performed by recording over time the photocurrent in potentiostatic mode at 0 V *vs.* RHE (*i.e.,* $J_{0V\ vs\ RHE}$) under continuous 1.5AM illumination.

Electrochemical impedance spectroscopy (EIS) is performed in H$_2$SO$_4$ solution at pH 1, at 0 V *vs.* RHE and under 1.5AM illumination. The spectra are recorded between 1 Hz and 100 kHz with AC amplitude of 10 mV.

## RESULTS AND DISCUSSION

### Architecture of the graphene-based organic photocathodes

The full structure of a photocathode based on solution-processed organic semiconductors comprises a transparent conductive substrate (*e.g.,* FTO and ITO), a HSL (*e.g.,* CuI, MoO$_3$, PEDOT:PSS), the photo-active layer, based on the polymer blend rr-P3HT:PCBM, an ESL (*e.g.,* TiO$_2$ and ZnO) and an EC layer (*e.g.,* Pt and MoS$_3$).[80,84,85] The operations of HSL/ESL concern:[163-166] 1) *charge extraction*, *i.e.,* the energy alignment between the conduction/valence bands of the HSL/ESL with respect of the highest occupied molecular orbital (HOMO)/LUMO levels of the organic active material, in order to create barrier-free potential and high-quality ohmic junctions that separate and inject the photogenerated charges;[163-166] 2) *charge selectivity*, depending by the relative position of the HSL CB (CB$_{HSL}$) with respect to the LUMO$_{PCBM}$ and that of the ESL VB (VB$_{ESL}$) with respect to the rr-P3HT HOMO (HOMO$_{P3HT}$) (*e.g.,* it is usually required that CB$_{HSL}$ > LUMO$_{PCBM}$ and VB$_{ESL}$ < HOMO$_{P3HT}$),[163-166] in order to reduce electron and hole recombination, respectively;[162-165] 3) *optical transparency* in the spectral range of absorption of the organic material, in order to avoid losses in the incident photonic flux;[163-165] 4) *surface smoothness,* in order to improve the quality of the contacts with the active layer;[163-166] 5) *PEC stability* in aqueous electrolytes in HER-working conditions.[82,83] Here solution-processed graphene-derivatives, *i.e.,* GO and RGO, are exploited as HSL candidates. Graphene oxide is synthesized from graphite flakes using modified Hummer's method,[167] while RGO is obtained by thermal annealing of the as-produced GO at 1000 °C[168] (see the Experimental Section, Synthesis of GO and RGO, for details). The morphology and electro-optical characterization of the as-produced GO and RGO is reported in Supporting Information -S.I.- (**Figure S1**, **Figure S2** and **Figure S3**). The structure of the photocathodes (*i.e.,* FTO/graphene-based HSL/rr-P3HT:PCBM/TiO$_2$/Pt-based EC) is obtained by depositing sequentially the material dispersions through low-temperature spin coating (see details in Experimental section, Fabrication of photocathodes).

**Figure 1**a shows the representative energy band edge positions of the photocathode materials together with the redox levels of the HER (-4.44 eV/0 V *vs.* vacuum level/HER) and OER (-5.67 eV/1.23 V *vs.* vacuum level/HER). The E$_F$ of GO (∼-4.9 eV) and RGO (∼-4.4 eV) are those determined by UPS and KP measurements (see S.I., Characterization of graphene-based materials, Figure S1d). The E$_F$ of GO shows better alignment with the HOMO$_{P3HT}$ level (∼-5 eV) if compared with the that of RGO. However, the metal-like behaviour of the RGO could, in principle, boost the holes transport towards the FTO.[163,164] Figure 1b shows the high-resolution cross-sectional SEM image of a representative photocathode, evidencing its multi-layered structure. TiO$_2$ and rr-P3HT layer are not observed as separated layers. This might be attributed to the partial penetration of the spin-coated TiO$_2$ in the rr-P3HT:PCBM underlay. However, the thickness of rr-P3HT:PCMB can be approximately estimated by that acquired in previous work on rr-P3HT:PCBM-based photocathodes.[77,83-85] The graphene-based HSL is not resolved in Figure 1b because of its very low (nanoscale) thickness value. Top-view SEM and elemental EDX analysis of FTO/GO and FTO/RGO is reported in S.I. (**Figure S4**, **Figure S5** and **Figure S6**). These results demonstrate that atomic-thick GO- and RGO-based HSLs are effectively deposited onto FTO (Figure S4).

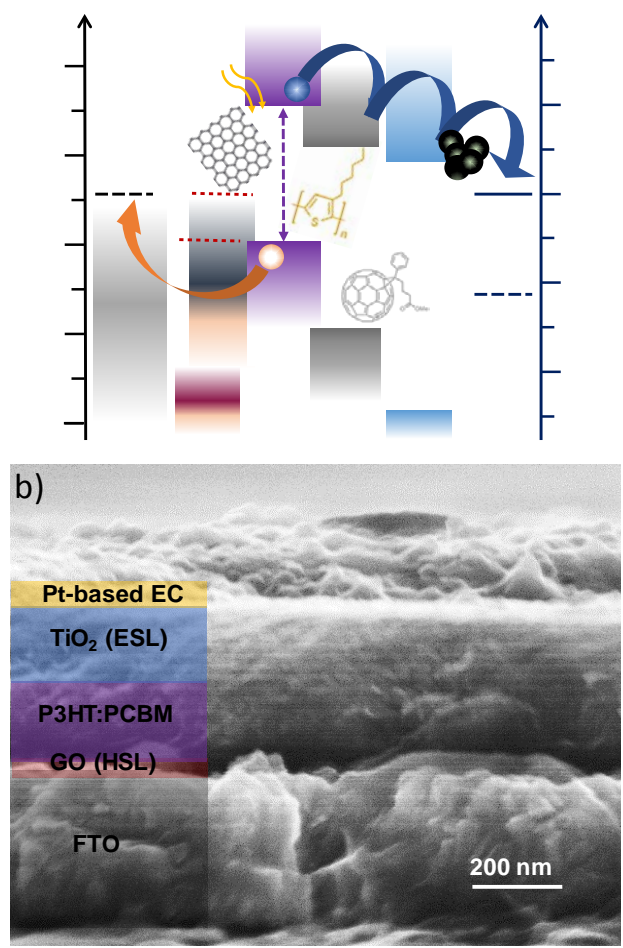

**Figure 1.** a) Scheme of the energy band edge position of the materials assembled in the solution-processed organic photocathode. The rr-P3HT:PCBM layer, in BHJ configuration, efficiently absorbs light and generates charges. The graphene-based layer and TiO$_2$ act as HSL and ESL, respectively, driving the holes towards the FTO substrates and the electrons towards the Pt nanoparticles, which act as EC layer for the HER process. Redox levels of both HER (blue solid line) and OER (blue dashed line) are shown. The E$_F$ of the GO (-4.9 eV) and RGO (-4.4 eV) are measured by UPS ambient KP measurements. b) High-resolution cross-sectional SEM image of the representative photocathode FTO/GO/rr-P3HT:PCBM/TiO$_2$/Pt-based EC.

The GO films does not affect the topography of the FTO (Figures S4-S6), while the formation of flakes aggregates hamper the homogeneity of the RGO films (Figure S4d). The aggregation of RGO flakes, as previously observed in rr-P3HT:PCBM based organic solar cells,[163,164,169] is attributed to the low dispersibility of RGO in polar solvents,[170-172] such as ethanol used here. This is a consequence of the limited content of oxygen functionalities (%c of C-O 6.9%) (see XPS analysis in S.I., Figure S1c), *i.e.*, loss of surface polarity, which determine a hydrophobic behaviour.[170-172] Thus, while GO dispersions are stable, we observed sedimentation of the RGO one as consequence of the poor hydrogen-bonding capability of the flakes (**Figure S7**).[170-172]

## Photoelectrochemical characterization

The rr-P3HT:PCBM-based photocathodes based on GO and RGO as HSLs, TiO$_2$ as ESL, and Pt nanoparticles as EC are characterized by LSV in H$_2$SO$_4$ solution at pH 1. Acid condition is initially chosen because HER kinetics on Pt are faster in acids than in neutral and alkaline solutions.[174-175] The LSVs of representative photocathodes based on GO and RGO deposited from dispersions at different concentration (0.5, 1 and 1.5 mg mL$^{-1}$) are reported in S.I., **Figure S8**. These results show that the best PEC performances are obtained for the dispersion at 1 mg mL$^{-1}$ for GO and 0.5 mg mL$^{-1}$ for RGO. The results obtained for representative photocathodes based on GO and RGO, as deposited from 1 and 0.5 mg mL$^{-1}$ dispersion in ethanol, respectively, are reported in **Figure 2**a. Their corresponding LSVs are compared with those of a HSL-free photocathode and the current-potential curve of Pt nanoparticles (*i.e.*, the EC) deposited directly onto the FTO. The LSVs of the photocathodes show a photocurrent that increases as the potential decreases. The photocurrents are positively affected by the presence of GO and RGO films, which are thus confirming their role of HSLs. The common FoM used to compare the performance of photocathodes are:[176] the $J_{0V\ vs\ RHE}$, the $V_o$, the $\Phi_{saved,NPA,C}$, and the $\Phi_{saved,ideal}$. In particular, $\Phi_{saved,NPA,C}$ and $\Phi_{saved,ideal}$ represent standard metrics to evaluate the overall performances of a photocathodes measured in a three-electrode configuration.[176] The definition of the FoM is reported in the Experimental Section, Photoelectrochemical characterization.

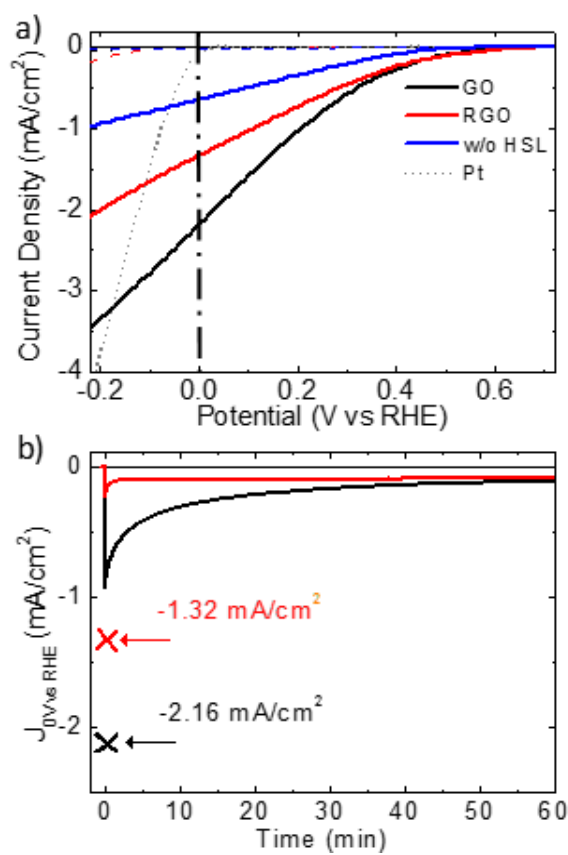

**Figure 2.** a) LSVs measured for the photocathodes using GO (black lines) and RGO (red lines) as HSLs measured in 0.5 M H$_2$SO$_4$ solution (pH 1), under dark (dashed lines) and AM1.5 illumination (100 mW cm$^{-2}$) (solid lines). GO and RGO films are deposited from 1 mg mL$^{-1}$ and 0.5 mg mL$^{-1}$ dispersions in ethanol. The PEC responses of the photocathode without any HSL (blue lines) and the current-potential curve of Pt nanoparticles (EC) deposited directly onto the FTO (short dashed grey line) are also shown. b) Potentiostatic stability tests of photocathodes using GO (black line) and RGO (red line), obtained by recording J$_{0V\ vs\ RHE}$ over 1 h of continuous AM1.5 illumination. The stability tests started

after the measurement of the LSVs shown in panel a). The values recorded in the LSVs of panel a) are also indicated in panel b).

The FoM of the photocathodes are $J_{0V\ vs\ RHE}$ = -2.16 mA cm$^{-2}$, $V_o$ = 0.56 V vs. RHE, $\Phi_{saved,NPA,C}$ = 0.29%, $\Phi_{saved,ideal}$ = 0.21% for GO and $J_{0V\ vs\ RHE}$ = -1.33 mA cm$^{-2}$, $V_o$ = 0.50 V vs. RHE, $\Phi_{saved,NPA,C}$ = 0.18%, $\Phi_{saved,ideal}$ = 0.15% for RGO. The better performances obtained by using GO with respect those recorded by using RGO are linked with the $E_F$ of GO matching ($\sim$-4.9 eV) with the HOMO$_{P3HT}$ ($\sim$-5 eV),[95,96] while the $E_F$ of RGO ($\sim$-4.4 eV) could lead to a rectifying (*i.e.*, Schottky barrier)[177] FTO/rr-P3HT contact for the hole extraction.[165]

Furthermore, the inhomogeneity of the RGO layer, evidenced by SEM analysis, Figures S4d, also affects the quality of the RGO/rr-P3HT junctions, determining charge recombination pathways (*i.e.*, leakage currents) in presence of blend-uncovered flake aggregates.[165,166] In order to assess the stability of our photocathodes in HER-working conditions, we carried out potentiostatic stability tests. These are performed by recording $J_{0V\ vs\ RHE}$ over 1 h continuous 1.5AM illumination (after LSV shown in Figure 2a). The results, reported in Figure 2b, show a performance degradation of the photocathodes. In fact, after the first LSV (where $J_{0V\ vs\ RHE}$ of -2.16 mA cm$^{-2}$ and -1.32 mA cm$^{-2}$ have been recorded for photocathodes using GO and RGO, respectively), $J_{0V\ vs\ RHE}$ at t = 0 is -0.93 mA cm$^{-2}$ for GO-based device and -0.23 mA cm$^{-2}$ for RGO-based one. No stabilization of the photocurrents towards constant values is observed, and after 1 h, $J_{0V\ vs\ RHE}$ decreases of ~95% and ~93% for GO- and RGO-based devices, respectively, with respect to the corresponding $J_{0V\ vs\ RHE}$ values in the LSV. The performances degradation can be caused by the detachment/dissolution of Pt from the TiO$_2$ surface, as previously reported for photocathodes in acid conditions.[67,83,178] Moreover, delamination/disruption of the layered structure of the photocathodes is macroscopically observed by eye (**Figure S9**a). These degradation effects are attributed to the poor adhesion between the different layers of the FTO/GO (RGO)/rr-P3HT:PCBM structure after the immersion in the electrolyte.[179,180] The delamination/disruption is instead not observed in our HSL-free photocathodes, in agreement with previous studies.[75-85]

## Stabilizing strategies

The decrease of the photocurrents observed during the potentiostatic stability test pointed out the need to implement stabilizing strategies to improve the endurance of the as-prepared photoelectrodes. Both Pt detachment/dissolution and delamination/disruption of the layered structure of the photocathodes based on GO and RGO can be at the origin of the performances degradation. In order to overcome these problems, two different stabilizing strategies are designed, as sketched in **Figure 3**. The first one relies on the fabrication of hydrogen-bonded FTO/graphene-based HSL/rr-P3HT:PCBM structures through the covalent linking between GO/RGO and a bi-functional silane compound, MPTMS (Figure 3a).[162]

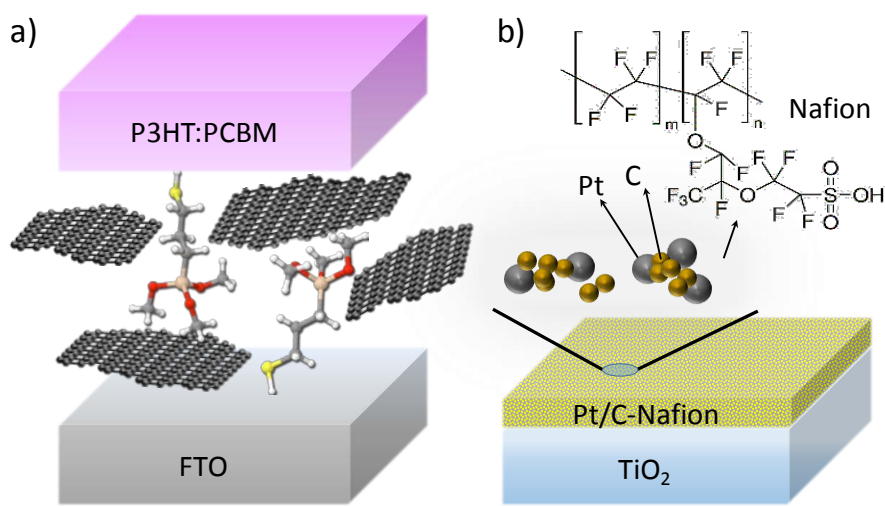

**Figure 3.** a) Silane-based chemical functionalization of GO/RGO for the fabrication of hydrogen-bonded FTO/graphene-based HSL/rr-P3HT:PCBM structures with improved mechanical adhesion properties. b) Implementation of solution-processed conductive and catalytic Pt/C-Nafion overlay for inhibiting the electrochemical degradation of the electrode materials and the multi-layered structures disruption.

The as-produced compounds (named f-GO and f-RGO, respectively) have silane groups anchored onto the f-GO and f-RGO flakes, while thiol groups (SH) are exposed to enhance the adhesion between adjacent layers of FTO/graphene-based HSL/rr-P3HT:PCBM structure. The second approach inhibits the electrochemical degradation of the electrode materials as well as the multi-layered structures delamination/disruption through the implementation of a solution-processed Pt/C-Nafion overlay (Figure 3b). The procedure of the silane-based functionalization of GO and RGO as well the formulation and the deposition of the Pt/C-Nafion overlay are reported in Experimental Section (Functionalization of GO and RGO and Fabrication of photocathodes, respectively).

The silane coupling with GO and RGO flakes is evaluated by means of XPS measurements, as reported in S.I., **Figure S10**. The morphology characterization of the functionalized materials and their corresponding films deposited onto the FTO is also reported in S.I., **Figure S11**, **Figure S12**, **Figure S13** and **Figure S14**. It is worth noting that, while RGO deposition determined the formation of large aggregates (Figure S4d), the deposition of f-RGO is not altering the characteristic morphology of the FTO (FTO grains are still visible on the high-magnification SEM image (Figure S12d). This is a consequence of the improved dispersion in ethanol in presence of MPTMS groups, which decreases the surface energy of RGO (~46.1 mN m$^{-1}$ in ethanol)[170,181,182] and enhance its compatibility with polar solvents such as ethanol.[170] As a consequence, there is an increase of the dispersion stability (Figure S7), avoiding the flakes aggregation during films deposition. In order to verify if the functionalization of the GO and RGO determined a change in the corresponding $E_F$, we carried out UPS analysis. The UPS data shows that the $E_F$ and the VB level of f-GO and f-RGO (**Figure S15**) are unchanged with respect to the ones of the starting GO and RGO flakes reported in Figure S1d.

The LSVs in $H_2SO_4$ solution at pH 1 of representative photocathodes based on f-GO and f-RGO deposited from dispersions at different concentration (0.5, 1 and 1.5 mg mL$^{-1}$) are reported in S.I. (**Figure S16**), showing that the best PEC performance for the dispersion at 0.5 mg mL$^{-1}$ for f-GO and 1 mg mL$^{-1}$ for f-RGO. The obtained values for the main FoM drastically decrease for the photocathodes based on f-GO (Figure S16a) ($J_{0V\ vs\ RHE}$ = -0.30 mA cm$^{-2}$, $V_o$ = 0.23 V *vs.* RHE and $\Phi_{saved,ideal}$ = 0.03%) with respect to the ones based on GO ($J_{0V\ vs\ RHE}$ = -2.16 mA cm$^{-2}$, $V_o$ = 0.50 V *vs.* RHE and $\Phi_{saved,ideal}$ = 0.21%). Different results are instead achieved with f-RGO (Figure S16b). In fact, a clear enhancement of the performance is observed for photocathodes based on f-RGO ($J_{0V\ vs\ RHE}$ = -1.82 mA cm$^{-2}$, $V_o$ = 0.5 V *vs.* RHE and $\Phi_{saved,ideal}$ = 0.19%) if compared with RGO-based ones ($J_{0V\ vs\ RHE}$ = -1.33 mA cm$^{-2}$, $V_o$ = 0.50 V *vs.* RHE and $\Phi_{saved,ideal}$ = 0.15%). The different FoM values achieved by photocathodes based on the functionalized materials could be due to the different mechanism for the hole extraction of GO and f-GO with respect to that of RGO and f-RGO.[163,164,168,183,184] From the $E_F$ and VB level estimated by UPS measurements (Figures S15), GO and f-GO are insulators, being able to extract the charge carriers through a quantum mechanical tunnelling process.[163,164,169,183,184] However, the presence of silane groups can alter the dipole formation between f-GO and rr-P3HT:PCBM,[185,186] thus varying the hole extraction processes.[183,184] Differently, RGO and f-RGO are metallic as deduced by UPS measurements (Figures S15) and can extract the charge carriers directly through their VB.[163,164,169,183,184] Here, the functionalization of the RGO flakes avoids the formation of aggregates, thus improving the quality of the contact between FTO/HTL and rr-P3HT. [163,164,169] The data of the main FoM obtained for the photocathodes based on GO, RGO, f-GO and f-RGO are summarized in Table 1. In addition, the potentiostatic stability measurements of the photocathode using f-RGO over 1 h of continuous AM1.5 illumination (Figure S16c) have shown a clear improvement in stability with respect to ones based on GO and RGO (Figure 2b). After the first LSV (where $J_{0V\ vs\ RHE}$ is -1.82 mA/cm$^2$), $J_{0V\ vs\ RHE}$ at t = 0 is -1.63 mA cm$^{-2}$, with a decrease of ~45% after 1 h operation, which however still provides a $J_{0V\ vs\ RHE}$ of ~-1 mA cm$^{-2}$. The improved $J_{0V\ vs\ RHE}$ (*i.e.,* $\Phi_{saved,NPA,C}$ and $\Phi_{saved,ideal}$) over time obtained by the f-RGO-based photocathodes with respect to the ones achieved by RGO and GO is linked with an enhancement of the mechanical stability of the electrode. In fact, delamination/disruption of the photocathodes, shown by the GO- and RGO-based photocathodes (Figure S9b), is not observed here. This result proves the beneficial role of the RGO flakes functionalization to strengthen the adhesion between the layers of the FTO/HSL/rr-P3HT:PCBM structure.

**Table 1.** FoM of photocathodes fabricated without HSL and using GO, RGO, f-GO and f-RGO (obtained by depositing their dispersion in ethanol at different concentration, *i.e.,* 0.5, 1 and 1.5 mg mL$^{-1}$) as HSL. The FoM values are obtained from the LSVs measured at pH 1.

| | Conc. (mg mL$^{-1}$) | $J_{OV\ vs\ RHE}$ (mA cm$^{-2}$) | $V_o$ (V vs RHE) | $\Phi_{saved,NPA,C}$ (%) | $\Phi_{saved,ideal}$ (%) |
|---|---|---|---|---|---|
| w/o HTL | - | -0.64 | 0.38 | 0.08 | 0.07 |
| GO | 0.5 | -0.56 | 0.39 | 0.07 | 0.06 |
| | 1.0 | -2.16 | 0.50 | 0.29 | 0.21 |
| | 1.5 | -1.35 | 0.48 | 0.22 | 0.17 |
| RGO | 0.5 | -1.33 | 0.50 | 0.18 | 0.15 |
| | 1.0 | -0.79 | 0.42 | 0.1 | 0.09 |
| | 1.5 | -0.11 | 0.03 | 0.01 | 0.01 |
| f-GO | 0.5 | -0.3 | 0.26 | 0.03 | 0.03 |
| | 1.0 | -0.14 | 0.10 | 0.02 | 0.01 |
| | 1.5 | -0.11 | 0.01 | 0.01 | 0.01 |
| f-RGO | 0.5 | -1.11 | 0.47 | 0.15 | 0.12 |
| | 1.0 | -1.82 | 0.50 | 0.25 | 0.19 |
| | 1.5 | -1.40 | 0.45 | 0.16 | 0.13 |

However, degradation of the photocathodes using f-RGO are still significant ($J_{OV\ vs\ RHE}$ loss of ~45% after 1 h of operation), indicating other causes of instability, such as TiO$_2$ reduction in acidic condition[187,188] and Pt detachment.[67,83,178] In order to further increase the photocathodes stability, we also designed solution-processed Pt/C-Nafion overlay (Figure 3b). Actually, in order to achieve durable and highly efficient photocathodes, the materials adopted for the protective overlay must be electrochemically stable in aqueous solution and, at the same time, sufficiently permeable to maintain the contact between the electrocatalytic Pt nanoparticles and the electrolyte,[72,71] allowing the photogenerated electrons to reach the Pt nanoparticles.[71] Furthermore, the processing conditions of the coating must be compatible with the underlying layers, easy scalable and cheap.[70] Recently, a solution-processed protective layer based-polymer of polyethyleneimine (PEI) has been proposed for rr-P3HT:PCBM-based photocathodes to prevent catalyst dislodging.[83] The implementation of this strategy permitted to achieve an operational activity of the photocathodes up to ~3 h, (~1 h without PEI protective layer).[83] Here, we focused on a different coating of the FTO/graphene-based HSL/rr-P3HT:PCBM/TiO$_2$ structures, based on solution-processed conductive and catalytic Pt/C-Nafion blend (Figure 3b). Based on the obtained PEC results discussed above, this protective methodology is applied only for the photocathodes using GO and f-RGO as HSL. The ratio of the materials used in the formulation of the Pt/C-Nafion overlay and the deposition parameters are reported in the Experimental Section, Fabrication of the photocathodes. Top-view SEM images of a representative photocathode are shown in **Figure 4**a and Figure 4b. The images evidence the presence of spherically shaped aggregates with a diameter <50 nm (see also **Figure S17**). Elemental EDX analysis (Figures 4c-g) is carried out in order to clarify the aggregates composition and the uniformity of the Pt/C-Nafion overlay. The C mapping reported in Figure 4e indicates that the observed aggregates are attributed to C nanoparticles, while Pt and Nafion, which are identified by the elemental mapping of Pt (Figure 4f) and F (Figure 4g) atoms, are homogeneously distributed over the TiO₂ layer.

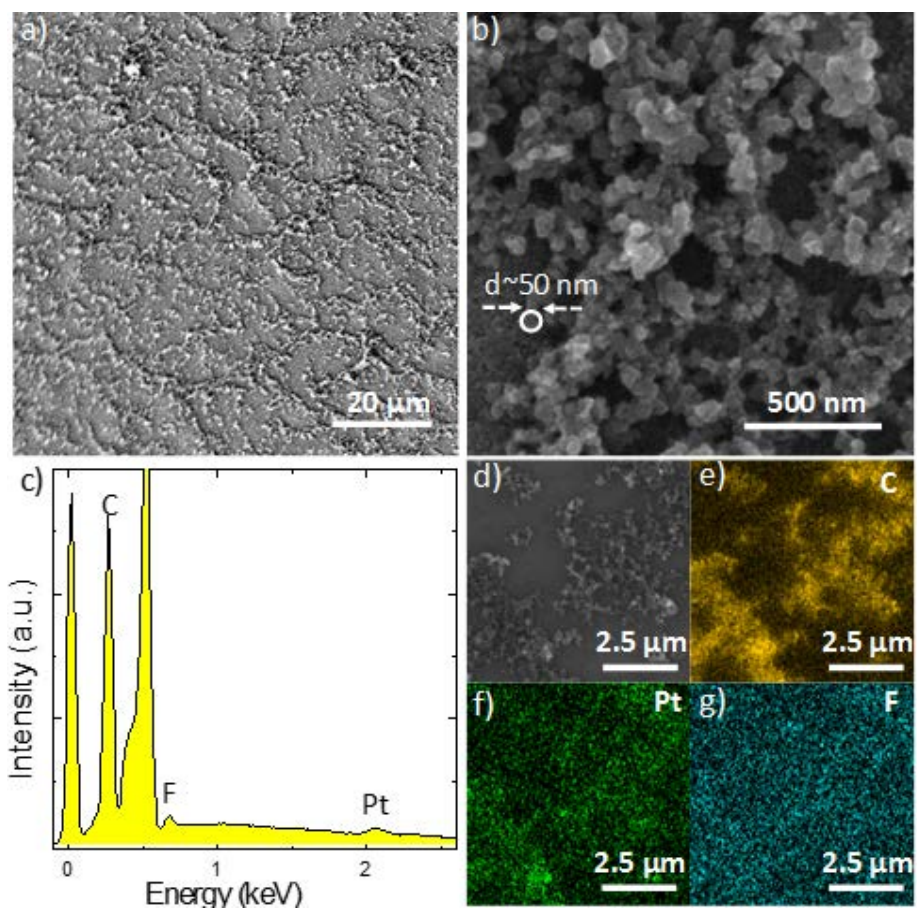

**Figure 4.** a) Top-view SEM (SEI) image of a Pt/C-Nafion layer covering a representative photocathode. b) Magnified SEM (SEI) image of panel a). c) Mass spectrum obtained by the EDX analysis of the images area shown in panel d). e) C, f) Pt and g) F mapping corresponding to the mass spectrum of panel c).

The PEC characterization of the photocathodes based on GO and f-RGO as HSL, and using Pt/C-Nafion overlay (named GO+Pt/C-Nafion and f-RGO+Pt/c Nafion, respectively) is reported in **Figure 5**a.

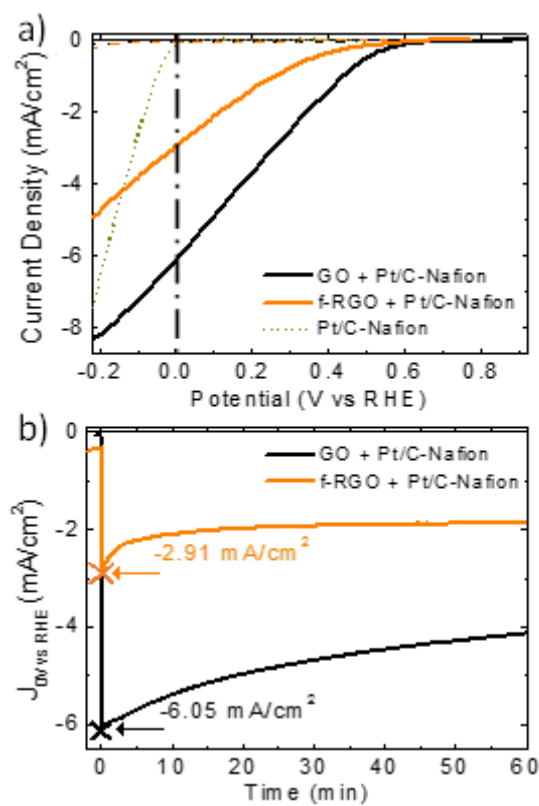

**Figure 5.** a) LSVs measured for the GO+Pt/C-Nafion (black lines) and f-RGO+Pt/C-Nafion (orange lines) as HSLs measured in 0.5 M $H_2SO_4$ solution (pH 1), under dark (dashed lines) and AM1.5 illumination (solid lines). GO and f-RGO films are deposited from 1 mg $mL^{-1}$ dispersions in ethanol. The current-potential curve of Pt/C-Nafion overlay deposited directly onto the FTO (short dashed dark yellow line) is also shown. b) Potentiostatic stability tests of GO+Pt/C-Nafion (black line) f-RGO+Pt/C-Nafion (orange line), obtained by recording $J_{0V\ vs\ RHE}$ over 1 h of continuous AM1.5 illumination. The stability tests started after the measurement of the LSVs shown in panel a). The values recorded in the LSVs of panel a) are also indicated in panel b.

The LSVs measurements show an improvement of the PEC performance of the photocathodes with respect to those without Pt/C-Nafion overlay (Figure S16 and Table 1). The main obtained results are $J_{0V\ vs\ RHE}$ = -6.01 mA $cm^{-2}$ (-2.93 mA $cm^{-2}$), $V_o$ = 0.60 V (0.55 V) *vs.* RHE, $\Phi_{saved,NPA,C}$ = 1.11% (0.36%), $\Phi_{saved,ideal}$ = 0.77% (0.27%), for the GO+Pt/C-Nafion (f-RGO+Pt/C-Nafion). The PEC performance achieved by the GO+Pt/C-Nafion-based photocathodes is remarkable. In fact, the measured efficiencies *(i.e., $\Phi_{saved,NPA,C}$* = 1.11% and $\Phi_{saved,ideal}$ = 0.77%), for FTO/GO/rr-P3HT:PCBM/$TiO_2$/Pt/C-Nafion architecture, are the current records measured of solution-processed rr-P3HT:PCBM photocathodes.[80,84,85] Moreover, the obtained results are approaching the current records measured for rr-P3HT:PCBM-based architecture produced through evaporation of protective metallic Ti-based layers[81] or pulsed laser deposition of $TiO_2$. However, the fabrication techniques for these photocathodes require controlled high vacuum[81] and/or atmosphere conditions,[83] and large scale uniform deposition,[189,190] which are challenging if compared with solution-processed techniques.[90-92,118-132]

Beside the aforementioned improvements in PEC performances, a clear increase of stability is also observed for the two photocathodes, as reported in Figure 5b. In fact, a remarkable $J_{0V \, vs \, RHE}$ = -4.14 mA cm$^{-2}$ for the case of GO, and -1.88 mA cm$^{-2}$ for the case of f-RGO is achieved, which correspond to a retention of 69% and 64%, respectively, after 1 h of continuous AM1.5 illumination.

## Photoelectrochemical responses at different pH

The development of photocathodes operating in neutral and alkaline conditions is crucial for their exploitation in tandem configuration systems.[106-110] In order to address this target, we tested our optimized photocathodes (*i.e.*, GO+Pt/c-Nafion and f-RGO+Pt/C-Nafion) at different pH, *i.e.,* acid, neutral as well alkaline conditions (**Figure 6**). The LSVs obtained at pH 1, 4, 7 and 10 for photocathodes using GO and f-RGO with Pt/C-Nafion overlay are reported in Figures 6a,b, respectively. Remarkable PEC activity is observed at all the pH conditions. For example, $J_{0V \, vs \, RHE}$ are -1.64 (-0.89), -1.51 (-0.91), -1.41 (-0.45) mA cm$^{-2}$ for GO+Pt/C-Nafion (f-RGO+Pt/C-Nafion) at pH 4, 7 and 10, respectively, are obtained.

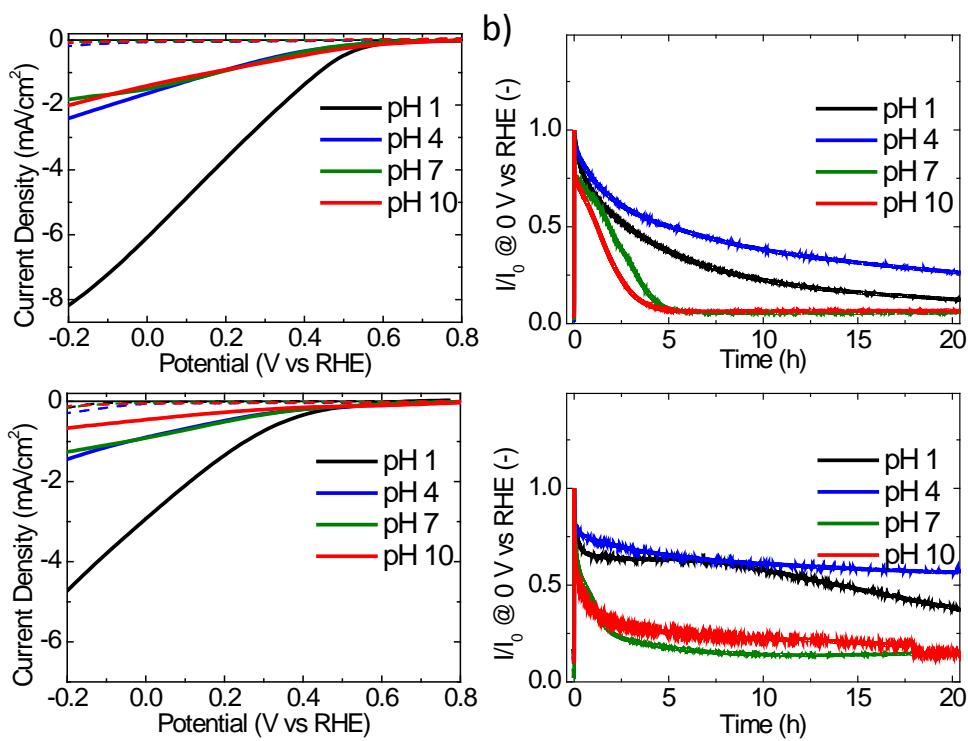

**Figure 6.** LSVs measured for the a) GO+Pt/C-Nafion and c) f-RGO+Pt/C-Nafion at pH 1 (black lines), 4 (blue lines), 7 (olive lines) and 10 (red lines) under dark (dashed lines) and AM1.5 illumination (solid lines). Potentiostatic stability tests b) GO+Pt/C-Nafion and d) f-RGO+Pt/C-Nafion, obtained by recording $J_{0V \, vs \, RHE}$ over 1 h of continuous AM1.5 illumination at pH 1 (black lines), 4 (blue lines), 7 (olive lines) and 10 (red lines). The photocurrents are normalized to the values of photocurrent at t = 0. The stability tests started after the measurement of the LSVs shown in panels a) and c).

Table 2 summarizes the values for the main FoM extracted from the voltammograms measured for the different pH, for both GO+Pt/C-Nafion and f-RGO+Pt/C-Nafion. Our results are the current record efficiencies, at the best of our knowledge, for rr-P3HT:PCBM-based photocathodes in neutral and alkaline conditions.[75]

**Table 2.** FoM of GO+Pt/C-Nafion and RGO+Pt/C/Nafion photocathodes. The FoM values are obtained from the LSVs measured at different pH (1, 4, 7 and 10).

| | pH | $J_{0V\ vs\ RHE}$ (mA cm$^{-2}$) | $V_0$ (V vs RHE) | $\Phi_{saved,NPA,C}$ (%) | $\Phi_{saved,ideal}$ (%) |
|---|---|---|---|---|---|
| GO +Pt/C- Nafion | 1 | -6.01 | 0.60 | 1.11 | 0.77 |
| | 4 | -1.64 | 0.55 | 0.23 | 0.19 |
| | 7 | -1.51 | 0.46 | 0.23 | 0.19 |
| | 10 | -1.41 | 0.60 | 0.23 | 0.20 |
| f-RGO +Pt/C- Nafion | 1 | -2.93 | 0.55 | 0.36 | 0.27 |
| | 4 | -0.89 | 0.56 | 0.11 | 0.10 |
| | 7 | -0.91 | 0.54 | 0.12 | 0.10 |
| | 10 | -0.45 | 0.60 | 0.06 | 0.06 |

Potentiostatic stability tests at different pH values for the GO+Pt/C-Nafion and f-RGO+Pt/C-Nafion are reported in Figures 6c,d, respectively. The data show better stability of the photocathodes operating at pH 1 and 4 with respect to pH 7 and 10. A retention of the $J_{0V\ vs\ RHE}$, with respect to its starting values, *i.e.,* 30% (64%) and 50% (66%) for pH 1 and 4, respectively, is measured for GO+Pt/C-Nafion (f-RGO+Pt/C-Nafion) after 5 h of continuous operation. After 20 h of endurance test for GO+Pt/C-Nafion (f-RGO+Pt/C-Nafion) shows a retention of the $J_{0V\ vs\ RHE}$, with respect to its starting values of 12% (38%) and 27% (57%) at pH 1 and 4, respectively. It is worth to note that our devices have shown a two-fold increase in durability if compared with state-of-the-art photocathodes based on WO$_3$ as HSL, tested at pH 1.37.[82] In fact, for the latter, a $\Phi_{saved,ideal}$ of ~0.25% was measured after 2 h of operation with a retention of the $J_{0V\ vs\ RHE}$ of 70% after 8 h.[82] At pH 7 and 10 photocurrents decrease rapidly during the first 5 h of operation. The $J_{0V\ vs\ RHE}$ decreases with respect to its starting values of 93% at both pH 7 and 10 for the GO+Pt/C-Nafion, and of 74% and 82% at pH 7 and 10, respectively, for the f-RGO+Pt/C-Nafion. The degradation here observed is attributed to the electrochemical instability of the Pt/C-Nafion overlay at neutral and basic conditions. This is evidenced by top-view SEM images of GO-Pt/C-Nafion photocathode before (**Figure S18a**) and after its immersion in the electrolyte at pH 10 (Figure S18b) and after 20 h of operation at 0 V *vs.* RHE and continuous AM1.5 illumination (Figure S18c). After the contact with the electrolyte, a clear re-distribution of the Pt/C network onto the TiO$_2$ surface is evidenced by the formation of Pt/C aggregates with larger dimensions (Figure S18b), if compared with the pristine ones (Figure S18a). After 20 h of operation, the surface is clearly damaged with no presence of the Pt/C (Figure S18c). This effect could proceed via a platinum dissolution/re-deposition mechanism or 3D Ostwald ripening[191] of the Pt/C-Nafion, due to C[192] and Pt[192,193] corrosion, which changes the adhesion of the materials between the Pt/C-Nafion overlay.[192] After the detachment/dissolution of the Pt/C-Nafion overlay, the underlying structure remains unprotected and exposed to the electrolyte, and the $H_2$ bubbling during HER causes a progressive "craterisation" of the photocathode surface, as evidenced in **Figure S19**.

## Flexible and large-area photocathodes

Photoelectrodes based on organic materials, such as graphene derivatives and photo-active conjugated polymers (*e.g.,* rr-P3HT), could in principles offer low manufacturing cost at high volume, thanks to their fast, low temperature, solution processing deposition on flexible plastic substrates.[90-92,115-132] Thus, we used graphene-based HSLs for fabricating large-area (9 cm$^2$) solution-processed rr-P3HT:PCBM-based photocathodes on flexible ITO-PET substrates (see fabrication details in Experimental section, Fabrication of photocathodes). **Figure 7**a and Figure 7b report the images of a representative solution-processed flexible 9 cm$^2$-area photocathode using GO as HSL and Pt/C-Nafion overlay (*i.e.*, GO+Pt/C-Nafion). Figure 7c shows the LSVs obtained for the large-area GO-based photocathodes as compared to those obtained for the corresponding 1 cm$^2$-area photocathode. The main FoM obtained for the 9 cm$^2$-area photocathode are $J_{0V\ vs\ RHE}$ = -2.80 mA cm$^{-2}$, $V_o$ = 0.45 V *vs.* RHE, $\Phi_{saved,NPA,C}$ = 0.31% and $\Phi_{saved,ideal}$ = 0.23%. It is important to note that the value obtained for $\Phi_{saved,NPA,C}$ approaches the previous literature records reported for solution-processed architecture on smaller scales (~1 cm$^2$-area) ($\Phi_{saved,NPA,C}$ = 0.47% and 0.43% for photocathodes using MoO$_x$[80] and MoS$_2$[84] as HSL, respectively, and MoS$_3$ as EC).[80,84] For the 9 cm$^2$-area device, the lower performances achieved with respect to the 1 cm$^2$-area one are attributed to the series resistance ($R_s$) of the photocathodes. The series resistance is given by the sum of the resistance of the substrate ($R_{FTO}$ or $R_{ITO}$), the resistance of the electrolyte ($R_{el}$) and the contact resistance ($R_c$). The values of $R_{FTO}$ and $R_{ITO}$ are equal to the sheet resistance ($R_{sh}$) of the substrates, which are ~15 Ω/□ for FTO and ~30 Ω/□ for ITO-PET substrates (here used as substrate for 1 cm$^2$-area and 9 cm$^2$-area photocathodes, respectively). Electrochemical impedance spectroscopy (EIS) measurements are carried out at 0 V *vs.* RHE and under 1.5AM illumination to evaluate the $R_s$ for the 1 cm$^2$-area and 9 cm$^2$-area photocathode. **Figure S20** reports the bode plots of the impedance (Z), *i.e.,* |Z| vs. frequency (f) (Figure S20a) and phase(Z) vs. f (Figure S20b), together with the corresponding Nyquist plots, *i.e.,* $Z_{im}$ vs. $Z_{re}$ (Figure S120c). The value of $R_s$ is estimated by the one of |Z| at high frequency (> 10$^4$ Hz).[194-196] From the EIS data, we obtained $R_s$ ~20 Ω for 1 cm$^2$-area and $R_s$ ~100 Ω for 9 cm$^2$-area one. The $R_s$ values of FTO (~20 Ω) is similar to that of its $R_{sh}$ (~15 Ω/□), thus excluding significant contribution from $R_{el}$ and $R_c$. Differently, the $R_s$ values of ITO-PET (~100 Ω) is remarkable higher than the nominal values of its $R_{sh}$ (~30 Ω/□). This is attributed to the slight shrinkage of the ITO/PET during the annealing process (*i.e.,* 130 °C for 10 min) in the fabrication of the photocathodes (see Experimental Section, Fabrication of the photocathodes), which leads to cracking of the ITO layer and consequently to the increase of its $R_{sh}$ value with respect to the nominal one.[197] The higher $R_s$ values observed for ITO-PET with respect to that of FTO causes the decrease of $V_{mpp}$ for the 9 cm$^2$-area photocathode ($V_{mpp}$ = 0.26 V *vs.* RHE) with respect to that of 1 cm$^2$-area one ($V_{mpp}$ = 0.17 V *vs.* RHE).[198,199] This also reflects the decrease of the FF from 0.21 in 1 cm²-area configuration to 0.16 in 9 cm²-area configuration, as illustrated in Figure 7d. It is worth to note that the $R_s$ of both FTO and ITO-PET substrates can be reduced by integrating metal grids onto ITO or FTO (*e.g.*, electroplated Cu grids)[198] or by connecting ITO or FTO through holes to a backside metallic electrode.[200] However, also by using these methods, uniform deposition of the overlays over large-area surface is still required for area-independent performance,[198] thus underlying the importance of the obtained results.

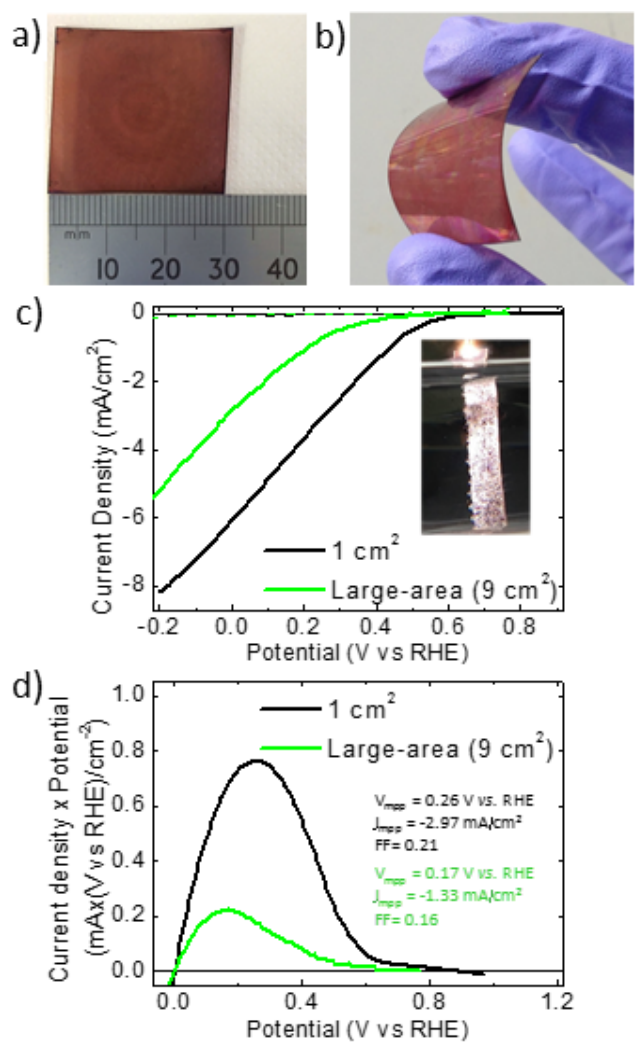

**Figure 7.** Photograph of a representative solution-processed large-area (9 cm²) ITO-PET/GO/rr-P3HT:PCBM/TiO₂/Pt/C-Nafion photocathode a) before and b) after bending. c) LSVs and d) current density × potential *vs.* potential curves measured for the GO+Pt/C-Nafion for 1 cm²-area and 9 cm²-area configurations (black and green lines, respectively) measured at pH 1 under dark (dashed lines) and AM1.5 illumination (solid lines). The inset to panel c) shows the hydrogen evolution on the surface of the photocathode operating at 0 V *vs.* RHE under 1.5AM illumination condition at pH 1. d) $V_{mpp}$, $J_{mpp}$, and FF, reported with the corresponding colours used for the LSVs, showing the decrease of FF by increasing the photocathode's area.

# CONCLUSION

Solution-processed GO and RGO atomic-thick films have been used as HSL to boost the efficiency and durability of rr-P3HT:PCBM-based photocathodes. By adopting silane-based functionalization of graphene derivatives-based HSLs, and a Pt/C-Nafion overlay, record-high performance and stability for solution-processed rr-P3HT:PCBM-based photocathodes are achieved at pH 1. Specifically, GO+Pt/C-Nafion photocathodes have shown $J_{0V\ vs\ RHE}$ = -6.01 mA cm$^{-2}$, $V_o$ = 0.6 V vs. RHE, $\Phi_{saved,ideal}$ = 1.11%, while f-RGO+Pt/C-Nafion ones reported $J_{0V\ vs\ RHE}$ = -2.93 mA cm$^{-2}$, $V_o$ = 0.55 V vs. RHE, $\Phi_{saved,ideal}$ = 0.27%. An operational activity of 20 h is reached at 0 V vs. RHE and under 1.5AM illumination condition. Moreover, the photocathodes are also effective at different pH values. The $\Phi_{saved,ideal}$ are 0.19%, 0.19% and 0.20% for GO+Pt/C-Nafion and 0.1%, 0.1% and 0.06% for f-RGO+Pt/C-Nafion, at pH 4, 7 and 10, respectively. We have furthermore demonstrated the up-scaling feasibility of our solution-processed devices, fabricating a flexible 9 cm$^2$-area photocathode achieving $J_{0V\ vs\ RHE}$ = -2.80 mA cm$^{-2}$, $V_o$ = 0.45 V vs. RHE, $\Phi_{saved,NPA,C}$ = 0.31%, $\Phi_{saved,ideal}$ = 0.23%.

In conclusion, our work demonstrates that organic photocathodes based on graphene derivatives represent an attracting technology to boost the commercialization of PEC devices for artificial photosynthesis.

## ACKNOWLEDGMENTS

This project has received funding from the European Union's Horizon 2020 research and innovation program under grant agreement No. 696656-GrapheneCore1. We thank the Electron Microscopy facility - Istituto Italiano di Tecnologia, for support in TEM data acquisition.

# Graphene-Based Hole Selective Layers for High-Efficiency, Solution-Processed, Large-Area, Flexible, Hydrogen-Evolving Organic Photocathodes

S. Bellani[a], L. Najafi[a], B. Martín-García[b], A. Ansaldo[a], Antonio E. Del Rio Castillo[a], M. Prato[c], I. Moreels[b] and F. Bonaccorso[1]*

[a] Graphene Labs, Istituto Italiano di Tecnologia, via Morego 30, 16163 Genova, Italy.
[b] Nanochemistry, Istituto Italiano di Tecnologia, via Morego 30, 16163 Genova, Italy.
[c] Materials Characterization Facility, Istituto Italiano di Tecnologia, via Morego 30, 16163 Genova, Italy.

‡ S. Bellani and L. Najafi contributed equally.

* Corresponding author: francesco.bonaccorso@iit.it.

## Characterization of as produced GO and RGO flakes

**Figure S1**a shows the UV-Vis absorption spectra of GO and RGO dispersions in ethanol. The GO spectrum reports a characteristic maximum at ~240 nm; this maximum is assigned to π→π* transition of C-C bonds,[1–3] and the broad shoulder between 290-300 nm is assigned as the π→π* transition of C=O bonds.[1–4] In the RGO spectrum, the maximum peak shifts to 275 nm and the absorption in the visible region increases, with respect to GO. This is linked with the restoration of the π-conjugation of $sp^2$ carbon atoms in the aromatic rings upon thermal reduction.[5–7] Moreover, the peak attributed to C=O bonds is significantly attenuated and red shifted of ~ 20 nm, with respect to the GO spectrum. This indicates the removal of oxygen-containing functional groups in the RGO.[5–7]

Structural changes of RGO with respect to the GO are also investigated by Raman spectroscopy, whose spectra are reported in Figure S1b. The Raman spectrum of GO reveals two main peaks located at 1352 and 1591 cm$^{-1}$, indicated as Pos(D) and Pos(G) corresponding to D and G bands, respectively.[8,9] The G peak corresponds to the $E_{2g}$ phonon at the Brillouin zone center,[8,9] while the D peak is due to the breathing modes of $sp^2$ rings,[8,9] requiring a defect for its activation by double resonance.[8] The 2D peak position (Pos(2D)), located at ~2700 cm$^{-1}$ is the second order of the D peak.[10] Double resonance can also happen as an intra-valley process, *i.e.,* connecting two points belonging to the same cone around K or K'.[10] This process gives rise to the D' peak, which is usually located at ~1600 cm$^{-1}$ in presence of high-density defects.[10] In these conditions, the D' band is merged with the G band. The 2D' peak, located at ~3200 cm$^{-1}$, is the second order of the D',[10] while D+D', positioned at ~2940 cm$^{-1}$ is the combination mode of D and D'. These three peaks show a low intensity, due to electronic scattering,[11] and a very broad lineshape.The full width half maximum (FWHM) of D (FWHM(D)) is 127 $cm^{-1}$, while FWHM(G) is 79 $cm^{-1}$. The FWHM(G) always increase with disorder and, indeed, it is much larger than pristine graphene (FWHM(G) < 20$cm^{-1}$)[9] and edge-defected graphene flakes (FWHM (G) ~ 25$cm^{-1}$).[12,13] The high intensity ratio between I(D) and I(G) ($I_D/I_G$) (~0.86) and the large FWHM(D) (~125$cm^{-1}$) is due to the presence of both structural defect (due to oxidation process) and covalent bonds (*e.g.*, C–H, C–O), both contributing to the D peak. In the case of RGO, Pos(D) is located at 1352 $cm^{-1}$, Pos(G) is at 1597 $cm^{-1}$, FWHM(G) is 64 $cm^{-1}$ and FWHM(D) is 83 $cm^{-1}$. The softening of the G band with respect to that of GO could be ascribed to the presence of defected regions as consequence of thermal stresses upon annealing.[14] FWHM(D) and FWHM(G) are narrower with respect to those of GO, indicating a restoration of the $sp^2$ rings.[8] The intensity ratio of the D and G bands ($I_D/I_G$) for RGO (~1.25) is considerable higher with respect to the GO one (~0.86). The $I_G$ is constant as a function of disorder because it is related to the relative motion of $sp^2$ carbons,[8] while an increase of $I_D$ is directly linked to the presence of $sp^2$ rings.[8,9] Thus, an increase of the $I_D/I_G$ ratio means the restoration of $sp^2$ rings.[8] Raman statistical analysis of the Pos(D), Pos(G), FWHM(D), FWHM(G) and the $I_D/I_G$ is reported in **Figure S2**.

The C atomic network and the associated oxygen functional groups in the GO and RGO are evaluated by X-ray photoelectron spectroscopy (XPS) measurements. Figure S1c shows that the C 1s spectrum of GO can be de-convoluted into four components:[15] the vacancies distorting the $sp^2$ network, the C-C bonds in the GO rings, the C-O groups, and the C=O groups, centred at (283.7±0.2), (284.7±0.2), (286.8±0.2) and (288.2±0.2) eV, respectively.[16–18] The corresponding atomic percentage contents (%c) show the prevalence of C-C (48.5%) and C-O bonds (41.5%). C=O bonds have still significant %c of 7.6%, while vacancies correspond to residual %c of 2.4%. These data indicate, as expected, the strong presence of the oxygen functionalities in GO.[19,20] Different results are obtained for the C 1s spectrum of RGO, which is clearly dominated by $sp^2$ C (75.6%, peak centred at 284.5 eV), while the C-O peak, centred at (286.9±0.3) eV, is strongly reduced (6.9%) with respect to same peak of GO. Vacancies-related and C=O bonds almost disappeared with respect to the GO case (indeed the reported fit is obtained with no vacancies and C=O contributions). Moreover, a residue of $sp^3$ C is still present (peak centred at 285 eV, %c = 8.5%) as well as carboxylate carbon O–C=O bonds, represented by the peak at 290.0 eV with a %c of 3.8%. These results indicate that the delocalized π-conjugated structure is almost fully restored in RGO.[21,22]

Ultraviolet photoelectron spectroscopy (UPS) analysis is performed to estimate the work function (WF), *i.e.,* the position of the Fermi energy ($E_F$) with respect to vacuum level, of GO and RGO flakes. Figure S1d shows the secondary electron cut-off (threshold) energies of the He-I (21.22 eV) UPS spectra of GO (~16.3 eV) and RGO (~16.7 eV). The corresponding WF values are 4.9 eV for GO and 4.4 eV for RGO. Consistent WF values (5 eV and 4.5 eV for GO and RGO, respectively) are obtained by ambient Kelvin probe (KP) measurements of GO and RGO films deposited onto FTO by spin coating 1 mg mL⁻¹ of their respective dispersions in ethanol (see details in the main text, Experimental Section, Fabrication of photocathodes).

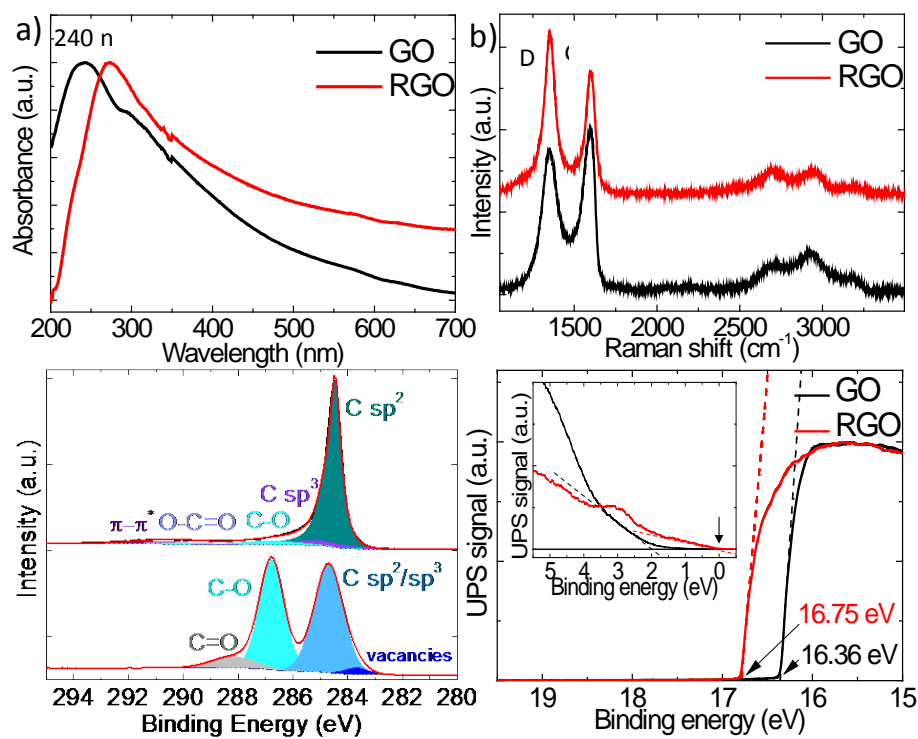

**Figure S1.** a) UV-Vis absorption spectra of GO (black line) and RGO (red line) dispersions in ethanol. The maximum absorption peaks (~240 nm for GO and ~275 nm for RGO), related to the π–π* transition of aromatic C–C bonds, are also evidenced. b) Raman of GO (black line) and RGO (red line) deposited onto a Si wafer with 300 nm thermally grown SiO₂. The main peaks G and D, the overtones 2D and 2D′ and the combination mode D+D′ are also evidenced, together with the ratio of D and G bands intensity $I_D/I_G$ (~0.86 for GO, and ~1.25 for RGO). c) C 1s spectra of GO and RGO. Their deconvolution is also shown. d) Secondary electron threshold region of He-I UPS spectra of GO (black line) and RGO (red line), which are used for estimating the WF values. The upper inset shows VB region of He-I UPS spectra of GO and RGO which are used for estimating WB values.

The higher WF of GO with respect to RGO is ascribed to the presence of surface dipole moments due to the oxygen functional groups, which disrupt the π-conjugation, as also evidenced by the XPS analysis (Figure S1c).[23–26] The upper inset of Figure S2d shows the spectra region near the $E_F$, which are used for estimating the valence band (VB) level of GO, ~-6.7 eV, and RGO, ~-4.4 eV (thus approaching its $E_F$ level). The relative distance between the VB and $E_F$ level of GO (~1.8 eV) indicates its insulating nature, while that of RGO (<0.1 eV) evidences its metal-like behaviour.[27,28]

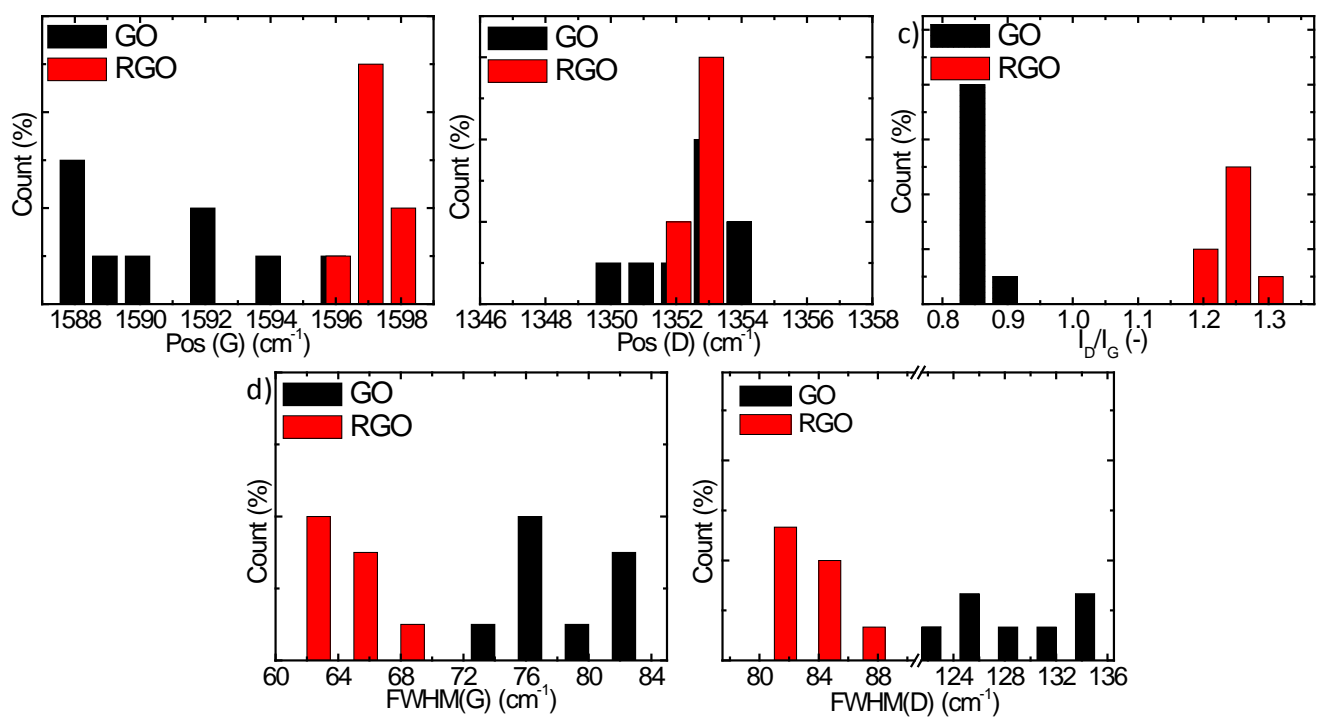

**Figure S2.** Statistical Raman analysis of the GO (black histograms) and RGO flakes (red histograms) for a) Pos(G), b) Pos(D), c) $I_D/I_G$, d) FWHM(G) and e) FWHM(D), calculated on 20 spots measured. The GO and RGO flakes are deposited from their respective ethanol dispersions onto a Si wafer with 300 nm thermally grown SiO$_2$.

The morphology (*i.e.*, lateral size and thickness) of the as-produced GO and RGO flakes is characterized by means of transmission electron microscopy (TEM) and atomic force microscopy (AFM). **Figure S3**a shows a representative TEM image of GO flakes, which have irregular shape and rippled morphology. Figure S3b shows the TEM image of RGO flakes, which have a more crumbled structure with respect to the GO ones. The TEM statistical analysis of the lateral size (Figure S3e) yields mean values of 2.8±1.6 μm for GO, and 1.7±0.8 μm for RGO. The changes of the RGO with respect to the GO are attributed to thermal-induced stress during the reduction treatment at high temperature (1000 °C).[29] Figure S3c,d show the AFM images of the GO and RGO flakes, respectively. Representative height profiles are also reported in Figure S3c,d (red lines), showing nano-edge steps between 0.6 and 1.6 nm indicating the overlap or multi-layered structures of the flakes. The AFM statistical analysis (Figure S3f) gives an average thickness of 1.7±0.9 nm and 1.8±1.1 nm for GO and RGO flakes, respectively. This indicates the few-layer nature of the as-produced flakes (thickness of single-layer pristine graphene is ~0.34 nm).[30,31]

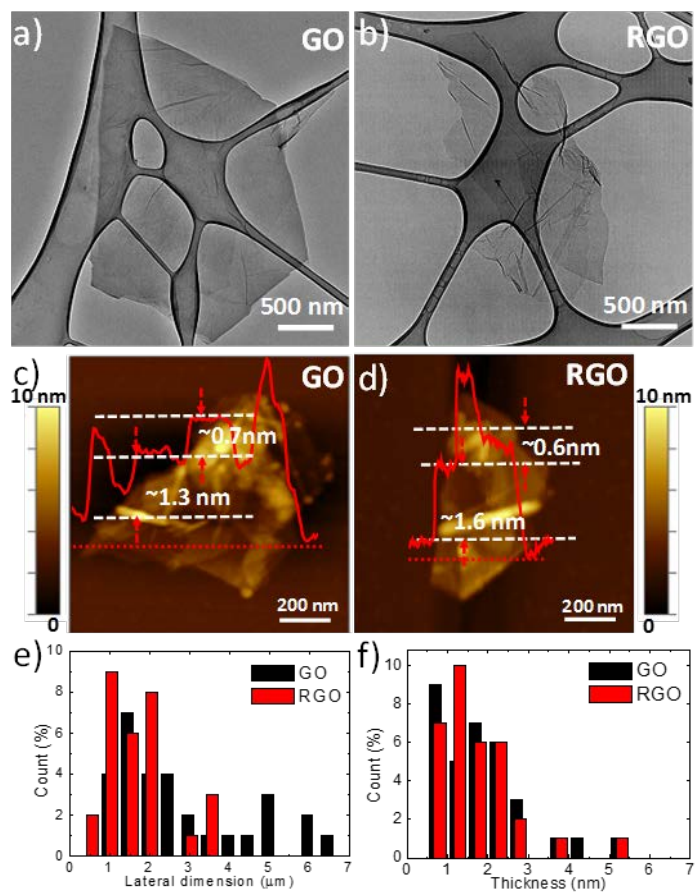

**Figure S3.** a) TEM images of the GO and b) RGO flakes drop casted onto carbon coated Cu TEM grids (300 mesh) from 0.01 mg mL$^{-1}$ dispersions in ethanol. c) AFM images of GO and d) RGO flakes deposited onto a V1-quality mica substrate from 0.1 mg mL$^{-1}$. Representative height profiles of representative flakes are also shown (red line). e) TEM statistical analysis of the lateral dimension of GO (black histograms) and RGO flakes (red histograms), derived from different images and calculated on 50 flakes. f) AFM statistical analysis of the thickness of GO flakes (black histograms) and RGO flakes (red histograms), derived from different images and calculated on 50 flakes.

## Top-view SEM analysis of FTO/GO and FTO/RGO

Top-view SEM images of FTO/GO (**Figure S4**a and Figure S4b) and FTO/RGO (Figures S4c,S4d) films provide a more detailed characterization of the HSL surface topography. Clearly, FTO crystal grains, as evidenced by SEM image of pristine FTO (**Figure S5**) are visible on a sub-µm scale (Figures S4a,c), indicating that no significant changes of the substrate surface occur after the GO and RGO deposition. Brighter regions, delimitated by red dashed lines, could be attributed to areas with low GO and RGO coverage. However, low-magnification SEM images (Figures 4b,d) evidence the presence of flakes aggregates. For the case of GO, the comparison between the chemical-sensitive SEM images collected using the in-lens sensor (upper secondary electron in-lens (SEI) image) (**Figure S6**a) with the topography-sensitive SEM images collected using the secondary electron sensor (lower secondary electron (LEI) image) (Figure S6b) indicates that the surface topography of FTO is not significantly affected by the presence of these aggregates, suggesting that they are nano-thick. On the other hand, for the case of RGO the substrate topography is clearly altered by the inhomogeneous film properties of the RGO (Figure S4d), which have been previously observed in rr-P3HT:PCBM based organic solar cells.[32-34] In order to confirm that atomic-thick HSLs effectively cover the FTO substrates, elemental energy-dispersive X-ray spectroscopy (EDX) analysis is performed on FTO/GO (Figures S4e-h) and on FTO/RGO (Figures S4i-n). Carbon atoms in the mass spectrum of Figures S4f,l are unambiguously attributed to the GO and RGO, and the C mapping in Figures S4g,m shows the homogeneity of C content onto the surface of FTO, identified by the Sn mapping (Figures S4h,n). In the case of RGO, an area with higher C content (delimited by dashed red line in Figure S4m) is ascribed to the presence of flakes aggregates, as evidenced by red dashed line in Figure 4i. The aggregation of RGO flakes is attributed to the low dispersibility of RGO in polar solvents,[35-37] such as ethanol used here. This is a consequence of the limited content of oxygen functionalities (%c of C-O 6.9%) (see previous XPS analysis, Figure S1c), *i.e.*, loss of surface polarity, which determine a hydrophobic behavior[35-37] Thus, while GO dispersions are stable, we observed sedimentation of RGO dispersion as consequence of the poor hydrogen-bonding capability of the flakes (see next section, Gravitational sedimentation of the GO, RGO, f-GO and f-RGO dispersion in ethanol).

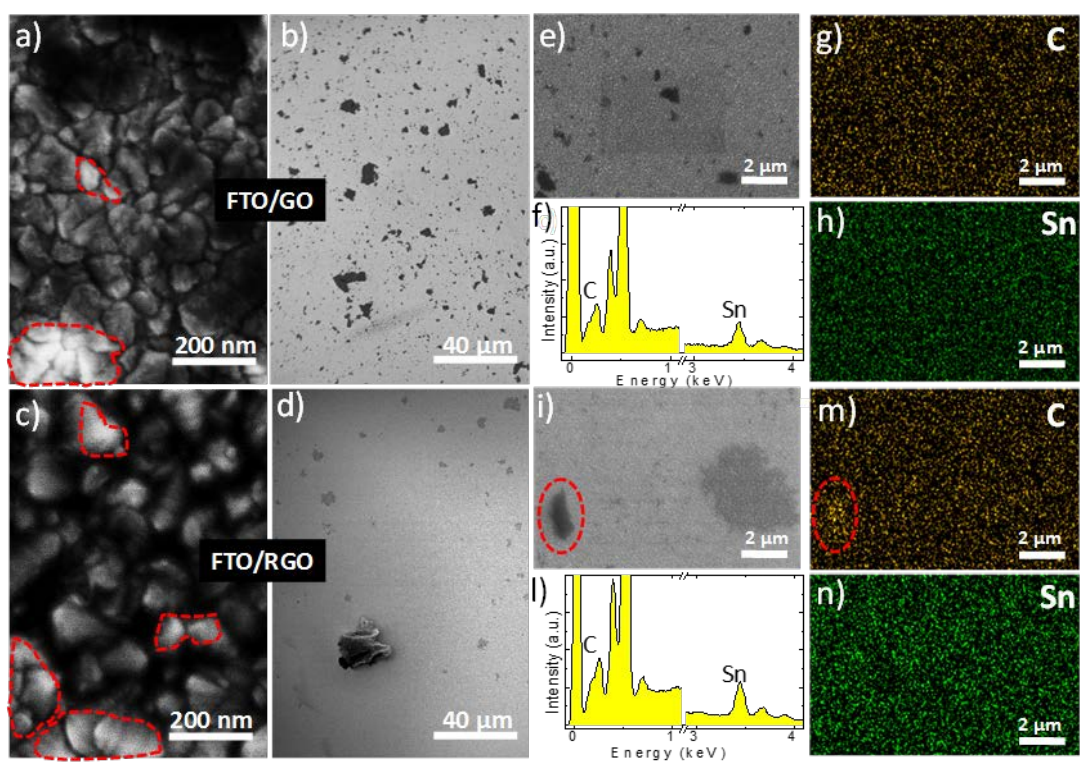

**Figure S4.** Top-view SEM images of a-b) GO and c-d) RGO film deposited on top of the FTO substrate from 1 mg mL$^{-1}$ ethanol dispersion. In a) and c) the bar scale is 200 nm, while b) and d) show a larger area (bar scale is 40 μm). e) Top-view SEM images of FTO/GO, on which elemental EDX analysis is performed. f) EDX spectrum of FTO/GO. g) C and h) Sn mapping corresponding to the EDX analysis of FTO/GO. i) Top-view SEM images of FTO/RGO, on which elemental EDX analysis is performed. l) EDX spectrum of FTO/RGO. m) C and n) Sn mapping corresponding to the EDX analysis of FTO/RGO. The areas delimitated by red dashed lines in a) and c)) indicate regions with lower level of GO and RGO coverage, respectively. The area delimitated by red dashed lines in i) and m) evidences the abundance of C due to the presence of GO and RGO flakes' aggregate.

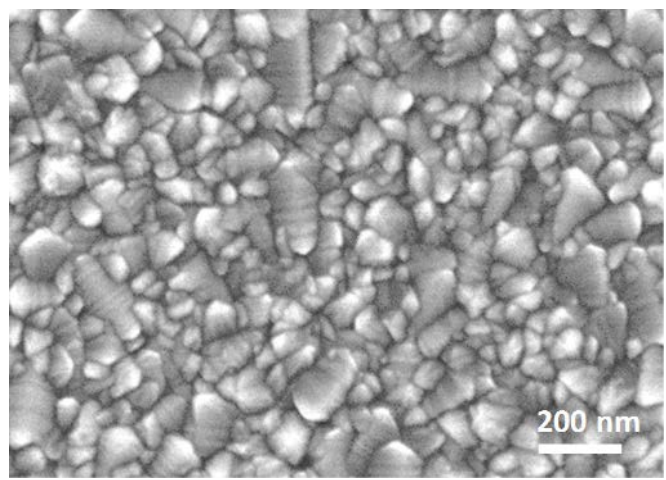

**Figure S5.** SEI-SEM images of FTO substrate.

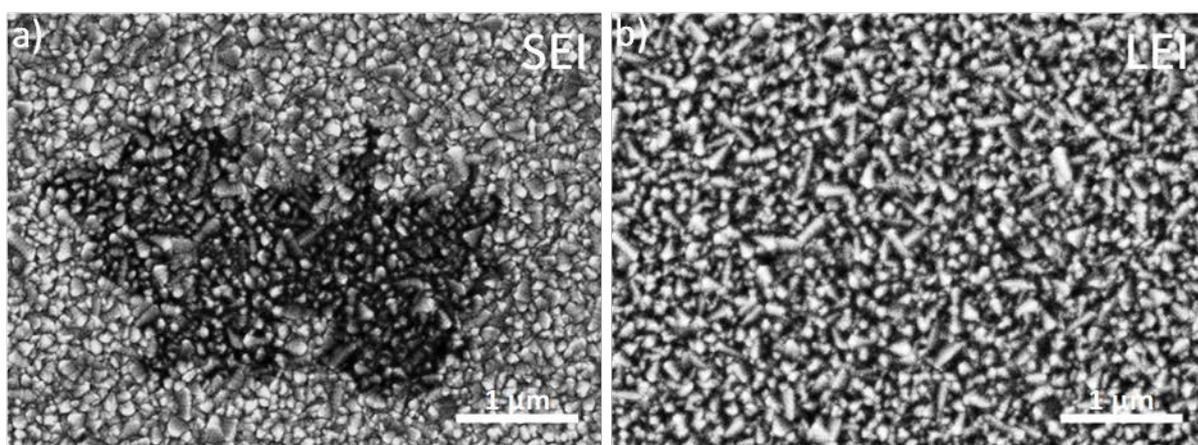

**Figure S6.** a) SEI-SEM and b) LEI-SEM images of FTO/GO sample. GO is deposited onto FTO by spin coating a 1 mg mL$^{-1}$ dispersion in ethanol.

## Gravitational sedimentation of the GO, RGO, f-GO and f-RGO dispersion in ethanol

**Figure S7** shows a photograph of 1 mg mL$^{-1}$ GO, RGO, f-GO and f-RGO dispersions in ethanol after 2 h of gravitational sedimentation. The photographs show a clear sedimentation of the RGO and GO dispersion as consequence of their poor hydrogen-bonding capability. After the silane functionalization of the materials, the presence of (3-mercaptopropyl)trimethoxysilane) MPTMS groups decreases their surface energy and enhances their compatibility with polar solvents such as ethanol, thus improving the stability of their dispersion. Consequently, no significant gravitational sedimentation for both f-GO and f-RGO dispersions in ethanol is observed.

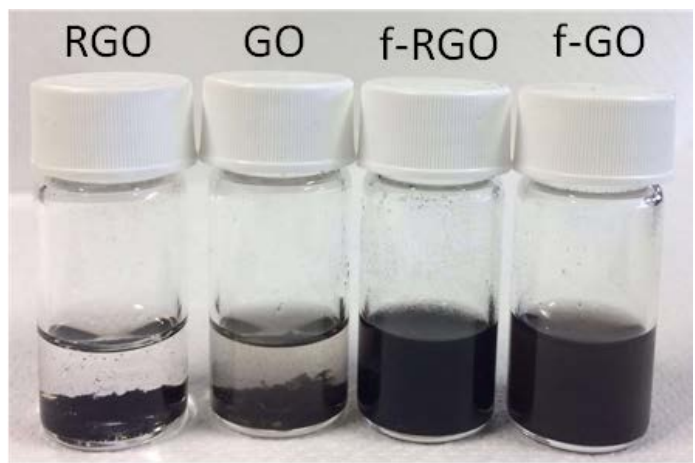

**Figure S7.** Photograph of 1 mg mL⁻¹ GO, RGO, f-GO and f-RGO dispersions in ethanol after 2 hour of gravitational sedimentation.

## Photoelectrochemical response of the photocathodes using GO and RGO deposited from different dispersion concentrations

**Figure S8** reports the linear sweep voltammetries (LSVs) of representative photocathodes using GO and RGO deposited from dispersions in ethanol at different concentration (0.5, 1 and 1.5 mg mL⁻¹), showing that the best (photo)electrochemical performances are obtained for the dispersion at 1 mg mL⁻¹ for GO and 0.5 mg mL⁻¹ for RGO. The data obtained for these last cases are reported in Figure 2a of the main text.

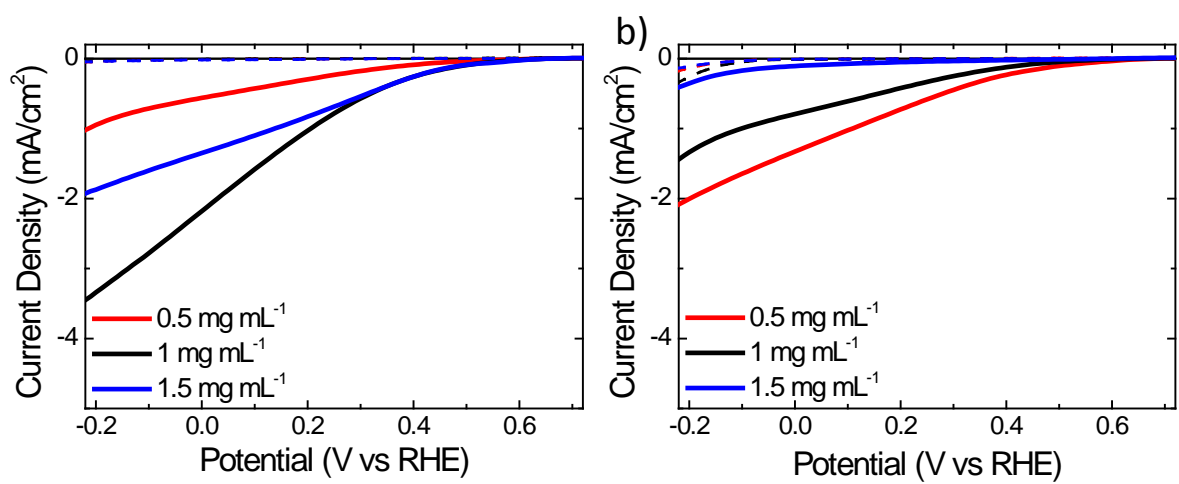

**Figure S8.** LSVs measured for the photocathodes using a) GO and b) RGO as HSLs deposited from dispersions in ethanol at different concentration: 0.5, 1 and 1.5 mg mL⁻¹ (red, black and blue lines, respectively), measured in 0.5 M H₂SO₄ solution (pH 1), under dark (dashed lines) and AM1.5 illumination (100 mW cm⁻²) (solid lines).

## Delamination/disruption effects on FTO/GO (RGO)/rr-P3HT:PCBM/TiO₂/Pt

For the photocathodes adopting the architecture FTO/(R)GO/rr-P3HT:PCBM/TiO$_2$/Pt, thus without the implementation of the stabilizing strategy discussed in the main manuscript, we carried out a potentiostatic stability test. The latter is performed by recording the photocurrents at 0 V *vs.* RHE (J$_{0V vs RHE}$) over 1 hour of continuous 1.5AM illumination, (see Fig. 2b in the main manuscript). After 1 h, the J$_{0V vs RHE}$ decreases of ~95% and ~93% for GO- and RGO-based devices, respectively, with respect to the corresponding J$_{0V vs RHE}$ values in the LSVs. The observed degradation is attributed to the poor adhesion between the different layer of the FTO/(R)GO/rr-P3HT:PCBM structure after the immersion in the electrolyte, as evidenced in **Figure S9**a. Notably, for the case of f-GO, no delamination/disruption effect is observed (Figure S9b).

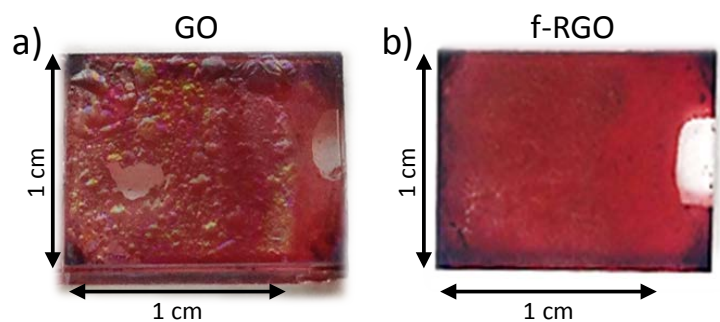

**Figure S9.** Photographs of a photocathode using a) GO and b) f-RGO as HSL after the potentiostatic stability test, obtained by recording J$_{0V vs RHE}$ over 1 hour of continuous 1.5AM illumination.

## Characterization of the f-GO and f-RGO flakes

The extent of silane functionalization of GO and RGO flakes is evaluated by means of XPS measurements. In the Si$_{2s}$ and S$_{2p}$ spectra (**Figure S10**a), the appearance of the silane- and thiol-related peaks at (153.4±0.3) eV and 163.4±0.3) eV, respectively,[38,39] indicates the effectiveness of the MPTMS functionalization procedure, although sulfur oxidation is observed (peaks around ~168 eV and ~169 eV related to S 2p doublet of SO$_4^{2-}$)[38,40,41] both in the f-GO and f-RGO samples. These oxidized S groups are either due to MPTMS molecules that interacted with oxygen moieties on the GO and RGO surfaces or, more likely, to a fraction of molecules that gets oxidized during the functionalization process itself. The oxidized S groups are the 33% and the 15% of the total S content for f-GO and f-RGO, respectively. The functionalization level is estimated from the ratio between the sum of the %c of SH free and S-S bonds related to the silane and that of C bonds (~0.06 and ~0.03 for f-GO and f-RGO, respectively). The functionalization level is estimated from the ratio between the %c of unoxidized S and that of C (~0.02 and ~0.01 for f-GO and f-RGO, respectively). The lower level of functionalization for f-RGO with respect to the one of f-GO is related to its low content of oxygen functionalities (Figure S1c), which act as anchor points for the silane groups[42–45] Moreover, the XPS analysis evidences an interconnection between the MPTMS molecules in the f-RGO case, since a low intensity S 2p doublet related to S-S bonds (centred at ~164.5 eV, accounting for 10% of the total S content) was needed for obtaining a good fit of the experimental data.[38,39] The increase of the C $sp^2$ %c of f-GO (58.2%) with respect to that of GO (48.5%) is attributed to the slight heating during the functionalization process. For the case of f-RGO flakes, the C $sp^2$ %c (75.7%) is the same observed for RGO flakes. Thus, the π-conjugated structure of RGO is not affected by the functionalization process.[21] In the meanwhile, the silane-based linkage of the f-GO/f-RGO flakes as well as the hydrogen-bonding capability of the free SH thiol groups anchored onto the f-GO and f-RGO flakes[46] enhance the interfacial adhesion between the layers of the FTO/graphene-based HSL/rr-P3HT:PCBM structure (Figure 3a).[47] The corresponding results (Figure S10) indicate a percentage functionalization level, as estimated from the ratio between the sum of the %c of SH free and S-S bonds related to the silane and that of C bonds, of ~6% and ~3% for f-GO and f-RGO, respectively (Figure S10a). Moreover, the π-conjugated structure of GO and RGO is not significantly affected by the functionalization process (Figure S10b).

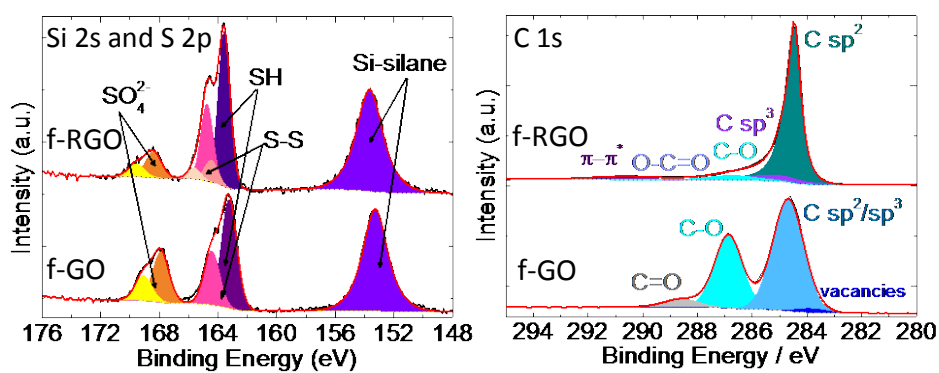

**Figure S10.** a) Si 2s and S 2p spectra of f-GO and f-RGO. Their deconvolution is also shown. b) C 1s spectra of GO and RGO. Their deconvolution is also shown.

The effect of chemical modification with silane functionalities on the morphology of the f-GO and f-RGO flakes, respect to the GO and RGO flakes, is investigated by means of transmission electron microscopy (TEM) and atomic force microscopy (AFM). **Figure S11**a and Figure S11b reports representative TEM images of f-GO and f-RGO flakes, respectively. For both cases, an irregular shape and rippled transparent paper-like morphology are observed, with a more crumbled structure for the f-RGO. The TEM statistical analysis of the flakes lateral dimensions (Figure S11e) reports mean values of ~2.8±1.4 μm for f-GO, and ~1.7±0.9 μm for f-RGO. Figures S11c,d show the AFM images of f-GO and f-RGO flakes, respectively, deposited onto a V1-quality mica. Representative height profiles are also reported in Figures S11c,d (red lines), showing nano-edge steps between 0.6 nm - 0.8 nm. This indicates the overlap or multi-layered structures of the flakes. The AFM statistical analysis (Figure S11f) gives an average thickness of 2.0±1.1 nm for f-GO flakes and 1.7±0.9 nm for f-RGO flakes. The values of lateral dimension and thickness obtained for f-GO and f-RGO are comparable with the ones of GO and RGO, respectively (Figure S3). Thus, the TEM and AFM results indicate that the chemical modification of the flakes, with silane functional groups, does not affect their native lateral dimension and thickness (mean values of 2.8±1.4 µm for f-GO (1.7±0.9 µm for f-RGO) and 2.0±1.1 nm for f-GO flakes (1.7±0.9 nm for f-RGO flakes), respectively).

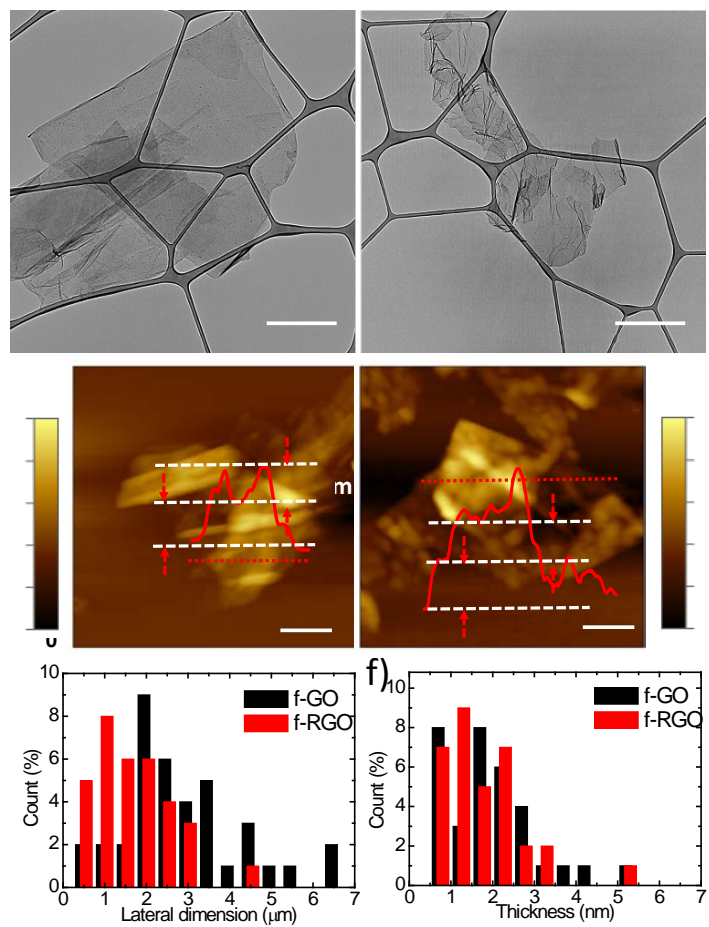

**Figure S11.** a) TEM images of the f-GO and b) f-RGO flakes drop casted onto carbon coated Cu TEM grids (300 mesh) from 0.01 mg mL$^{-1}$ dispersions in ethanol. c) AFM images of f-GO and d) f-RGO flakes deposited onto a V1-quality mica substrate from 0.1 mg mL$^{-1}$. Representative height profiles of representative flakes are also shown (red line). e) TEM statistical analysis of the lateral dimensions of GO (black histograms) and RGO flakes (red histograms), obtained from different images and calculated on 50 flakes. f) AFM statistical analysis of the thickness of GO flakes (black histograms) and RGO flakes (red histograms), obtained from different images and calculated on 50 flakes.

High-resolution TEM images of f-GO (**Figure S12**a) and f-RGO (Figure S12b) show darker grey spots (different contrast with respect to the GO and RGO flakes) that are attributed to the presence of MPTMS molecules anchored over the flakes. Top-view SEM images of FTO/f-GO (Figure S12c) and FTO/f-RGO (Figure S12d) show that no significant changes of the FTO substrate surface occurred after f-GO or f-RGO deposition (FTO grains are still visible under the flakes).

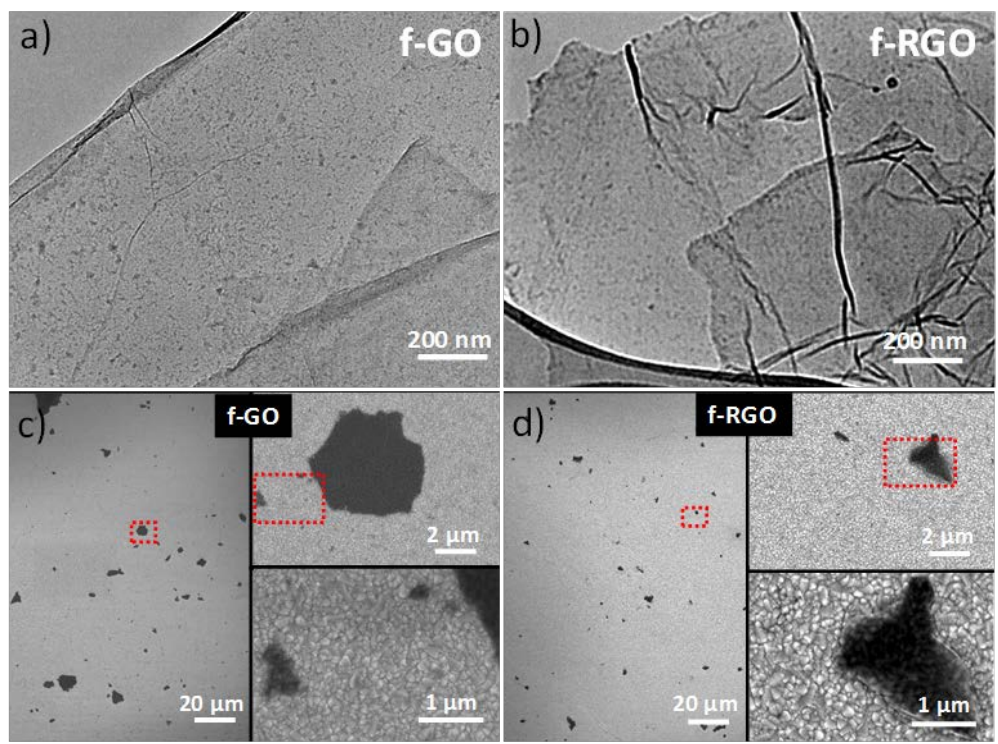

**Figure S12**. a) High-resolution TEM imageg of f-GO and b) f-RGO casted onto carbon coated Cu TEM grids (300 mesh) from 0.01 mg mL$^{-1}$ dispersions in ethanol. b) Top-view SEM images of f-GO layer deposited atop the FTO substrate from a 1 mg mL$^{-1}$ ethanol solution. Three panels show images with different magnification (20 μm, 2 μm and 1 μm bar scales) d) Top-view SEM images of f-RGO layer deposited on top of the FTO substrate from a 1 mg mL$^{-1}$ ethanol solution. Three panels show images with different magnification (20 μm, 2 μm and 1 μm bar scales). The areas delimited by red dashed lines in e) and f) indicate regions shown by images with higher magnification.

Elemental EDX analysis of FTO/f-GO (**Figure S13**) and FTO/f-RGO (**Figure S14**) indicates that C and Si atoms, which are attributed to the f-GO and f-RGO flakes, are homogeneously distributed over the FTO, as already shown for GO and RGO layers (Figure S4). It is worth noting that, while RGO deposition determined the formation of large aggregates (Figure S4d), the deposition of f-RGO is not altering the characteristic morphology of the FTO (FTO grains are still visible on the high-magnification image (Figure S12d). This is a consequence of the improved dispersion in ethanol in presence of MPTMS groups, which decreases the surface energy of RGO (~46.1 mN m$^{-1}$ in ethanol)[35,48,49] and enhance its compatibility with polar solvents such as ethanol,[35] increasing the dispersion stability, see gravitational sedimentation test in Figure S7, and avoiding the flakes aggregation during film deposition.

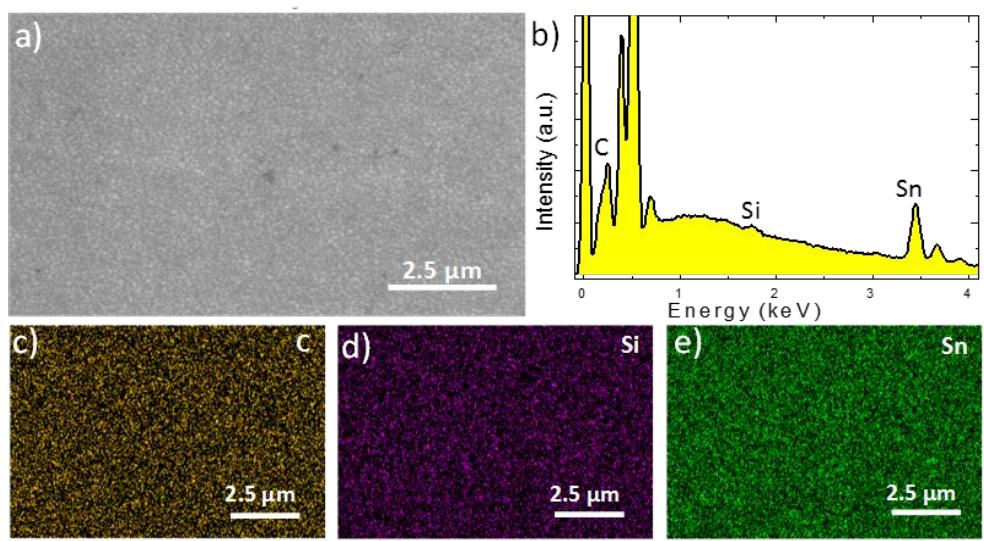

**Figure S13.** a) Top-view SEI-SEM image of FTO/f-GO, where elemental EDX analysis is performed. b) Mass spectrum of the EDX analysis. c) C, d) Si and e) Sn mapping obtained from the EDX analysis.

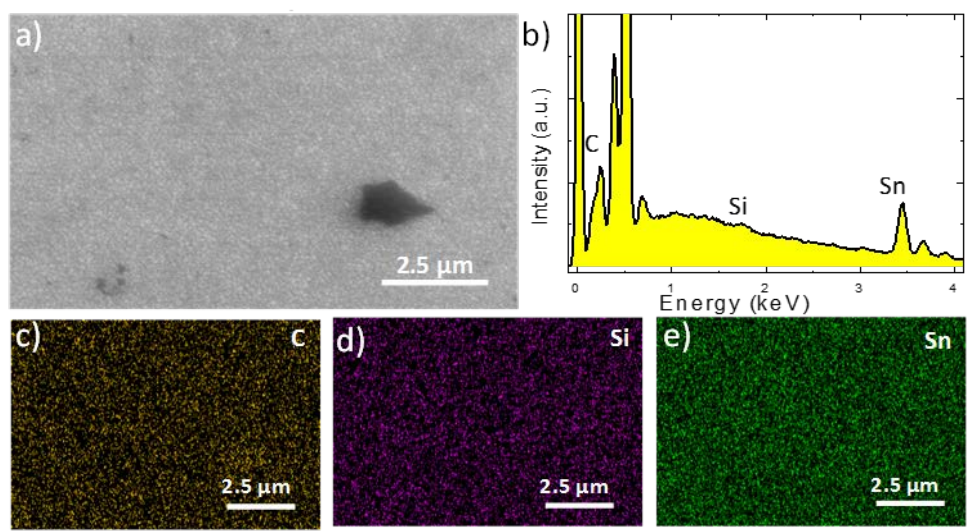

**Figure S14.** a) Top-view SEI-SEM image of FTO/f-RGO, where elemental EDX analysis is performed. b) Mass spectrum of the EDX analysis. c) C, d) Si and e) Sn mapping resulted from the EDX analysis.

The UPS analysis of f-GO and f-RGO is reported in **Figure S15**. The UPS measurements on GO and RGO are also reported for comparison. The secondary electron cut-off energies of the He-I (21.22 eV) UPS spectra, as obtained by applying a bias of -9 V to the samples, of GO and f-GO is equal (~16.3 eV), while a slight difference of 0.1 eV is observed for RGO (~16.7 eV) with respect to f-RGO (~16.8 eV). The corresponding energy Fermi ($E_F$) level are -4.9 eV for GO and f-GO, -4.4 eV for RGO and -4.3 eV for f-RGO. Consistent $E_F$ level (-5 eV for GO and f-GO, -4.5 eV for RGO and f-RGO) are obtained by ambient kelvin probe (KP) measurements of few-nm films deposited on FTO (see details in Experimental Section, Fabrication of photocathodes in the main text of the manuscript). The deeper $E_F$ of GO and f-GO with respect to that of RGO and f-RGO is attributed the presence of surface dipole moments attributable to the oxygen functional groups which disrupt the π-conjugation, as evidenced by the XPS analysis (Figures S1c,S10b). Importantly, the functionalization with silane molecules does not introduce significant changes. The inset of Figure S15 shows the spectra region near $E_F$, which are used for estimating the VB level of the materials (∼-6.7 eV for GO, ∼-6.9 eV for f-GO, and -4.4 for both RGO and f-RGO). These results indicate that the insulating behaviour of GO is also confirmed for the f-GO, while the metal-like behaviour of RGO is still preserved by the f-RGO.

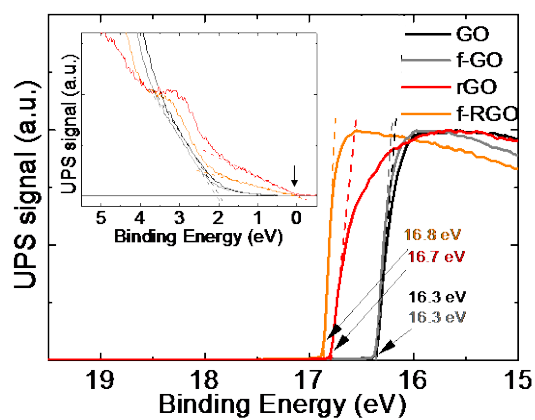

**Figure S15.** Secondary electron threshold region of He-I UPS spectra of GO (black line), f-GO (grey line), RGO (red line) and f-RGO (orange line). The upper inset shows VB region of He-I UPS spectra of GO and RGO which are used for estimating VB values.

## Photoelectrochemical response of the photocathodes using f-GO and f-RGO deposited from different dispersion concentrations

**Figures S16**a and Figure S16b report the LSVs of representative photocathodes using f-GO and f-RGO, respectively, deposited from the corresponding dispersions in ethanol at different concentration (0.5, 1 and 1.5 mg mL$^{-1}$). The data show that the best photoelectrochemical performances are obtained for the dispersion at 0.5 mg mL$^{-1}$ for f-GO and 1 mg mL$^{-1}$ for f-RGO. The potentiostatic stability measurements of the photocathode using f-RGO over 1 hour of continuous AM1.5 illumination (Figure S16c) show a clear improvement in stability with respect that of photocathode using GO and RGO. After the first LSV scan (where $J_{0V\,vs\,RHE}$ is -1.82 mA cm$^{-2}$), $J_{0V\,vs\,RHE}$ at t = 0 is -1.63 mA cm$^{-2}$. After 1 h of operation, $J_{0V\,vs\,RHE}$ decreases of ∼45% with respect to the $J_{0V\,vs\,RHE}$ value in the first LSV. However, the f-RGO-based device still provides a $J_{0V\,vs\,RHE}$ of ∼-1 mA cm$^{-2}$. The improved $J_{0V\,vs\,RHE}$ over time obtained by the f-RGO-based photocathodes with respect to the ones achieved by RGO and GO is linked with an enhancement of the mechanical stability of the electrode. Delamination/disruption of the photocathodes is not observed, proving the mechanical adhesion between the layers of the FTO/f-RGO/rr-P3HT:PCBM structure.

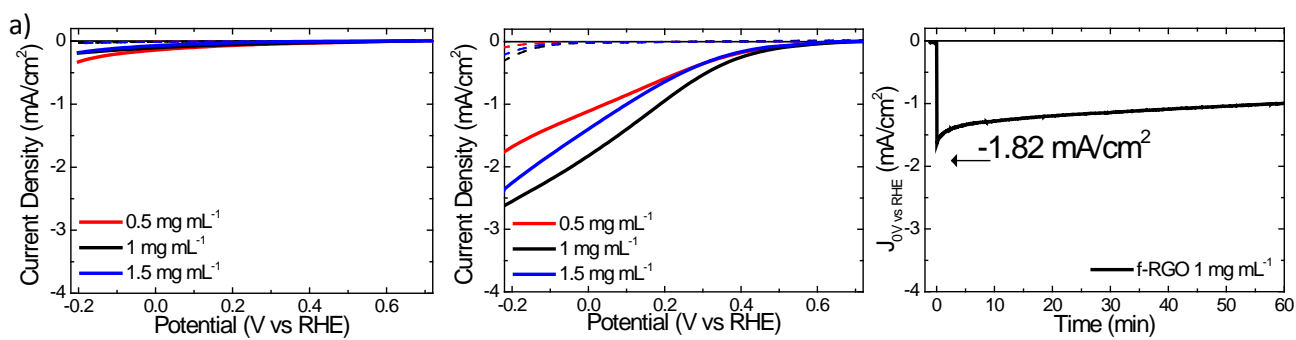

**Figure S16.** LSVs measured for the photocathodes using a) f-GO and b) f-RGO as HSLs deposited from dispersions in ethanol at different concentration: 0.5, 1 and 1.5 mg mL$^{-1}$ (red, black and blue lines), measured in 0.5 M H$_2$SO$_4$ solution (pH 1), under dark (dashed lines) and AM1.5 illumination (100 mW cm$^{-2}$) (solid lines). c) Potentiostatic stability test of photocathode using f-RGO obtained by recording J$_{0V}$ $_{vs RHE}$ over 1 h of continuous AM1.5 illumination.

## Top-view SEM image of a Pt/C-Nafion electrocatalytic overlay

**Figure S17** reports the top-view SEM image of a Pt/C-Nafion electrocatalytic layer for a representative photocathode. The image has a higher magnification with respect to those of the images reported in the main text (Figure 4a,b). The image evidences the presence of spherically shaped aggregates with a diameter smaller than 50 nm, which are attributed to carbon nanoparticles (see EDX analysis in Figure 4c-g of the main text).

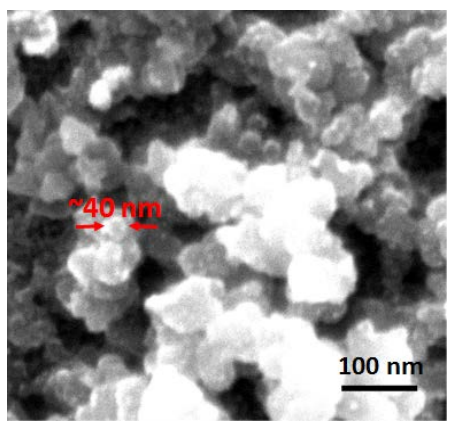

**Figure S17.** Top-view SEI-SEM image of a Pt/C-Nafion layer covering a representative photocathode. Red arrows evidence the dimension of a representative nanoparticle.

## Instability of Pt/C-Nafion electrocatalytic overlay at pH 10

**Figure S18** shows top-view SEM images of GO-Pt/C-Nafion photocathode before (Figure S18a) and after its immersion in the electrolyte at pH 10 (see Figure S18b), and after 20 h of operation at 0 V *vs.* RHE and continuous AM1.5 illumination condition (Figure S18c). After contact with the electrolyte, a clear redistribution of the Pt/C network is observed by the formation of more dispersed Pt/C aggregates on top of the surface (Figure S18b), with respect to those in the post-fabrication condition (Figure S18a). After 20-h operation the surface is clearly damaged and Pt/C network is not present anymore (Fig. S18c). These effects could proceed via a Pt dissolution/re-deposition mechanism or 3D Ostwald ripening[50] of the Pt/C-Nafion, because of both C[51] and Pt[51,52] corrosion. The latter changes the adhesion of the Pt/C-Nafion overlay.[52] After the detachment/dissolution of the Pt/C-Nafion overlay, the underlying structure remains unprotected, and the hydrogen bubbling during HER causes a progressive "craterisation" of the surface. All these effects are evidenced in **Figure S19**, especially the "craterisation" and the consequent exposure of the FTO substrate to the electrolyte.

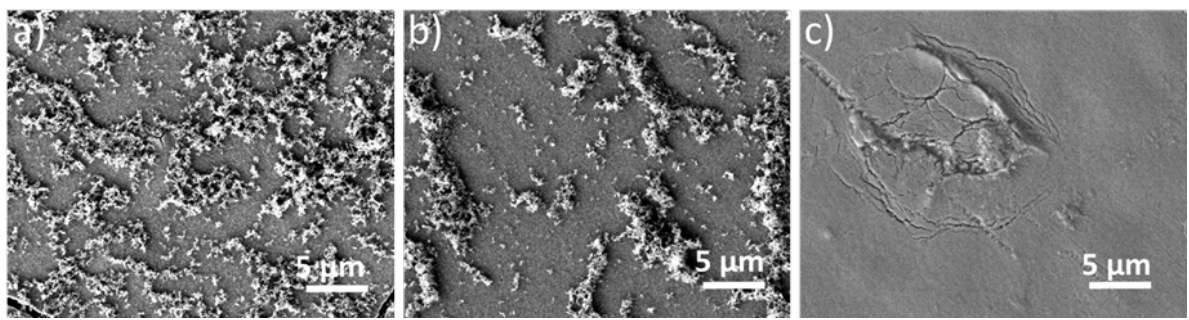

**Figure S18.** Top-view SEM images of the GO+Pt/C-Nafion a) immediately after its fabrication, b) after its immersion in the aqueous solution at pH 10 and c) after 20 hours of operation at 0 V *vs.* RHE and continuous AM1.5 illumination in the same aqueous solution at pH 10.

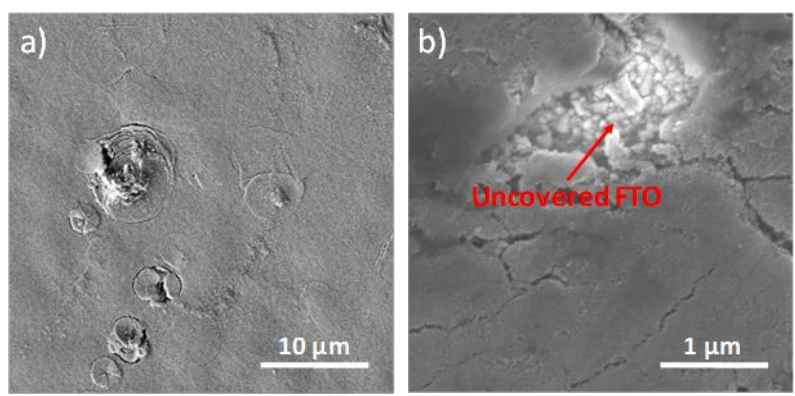

**Figure S19.** a) Top-view SEM images of the GO+Pt/C-Nafion after 20 hours of operation at 0 V *vs.* RHE and continuous AM1.5 illumination at pH 10. b) Top-view SEM images of the same photocathode focusing on a damaged area, where delamination/disruption of the device's structure is evidenced by the presence of uncovered regions of FTO.

## Electrochemical impedance spectroscopy

Electrochemical impedance spectroscopy (EIS) measurements are carried out at 0 V *vs.* RHE and under 1.5AM illumination to evaluate the series resistance ($R_s$) for the 1 cm$^2$ (FTO/GO/rr-P3HT:PCBM/TiO$_2$/Pt/C-Nafion) and 9 cm$^2$-area photocathode (ITO/GO/rr-P3HT:PCBM/TiO$_2$/Pt/C-Nafion). **Figure S20** reports the bode plots of the impedance (Z), *i.e.,* |Z| *vs.* frequency (f) (Figure S20a) and phase(Z) vs. f (Figure S20b) and the corresponding Nyquist plot, *i.e.,* $Z_{im}$ vs. $Z_{re}$ (Figure S20c). As discussed in the main text, $R_s$ ~20 Ω for 1 cm$^2$ and $R_s$ ~100 Ω for 9 cm$^2$ sample are estimated by |Z| value at high frequency (10$^4$ Hz). While the $R_s$ value of FTO/glass (~20 Ω) is similar to that of its sheet resistance ($R_{sh}$) (~15 Ω/□), thus excluding significant contribution from $R_{el}$ and $R_c$, the $R_s$ one of ITO-coated poly(ethylene terephthalate) (ITO-PET) (~100 Ω) is remarkable higher with respect to the nominal value of its $R_{sh}$ (~30 Ω/□). The different resistive behaviour of the $R_s$ for ITO-PET with respect that of FTO deposited on glass is ascribed to the slight shrinkage of the ITO-PET during the annealing process in the fabrication of the photocathodes (130 °C for 10 min) (see Experimental, Fabrication of the photocathodes). In fact, this phenomena cause the partial cracking of the ITO layer, thus the consequent increase of its $R_{sh}$ value with respect to the nominal one.

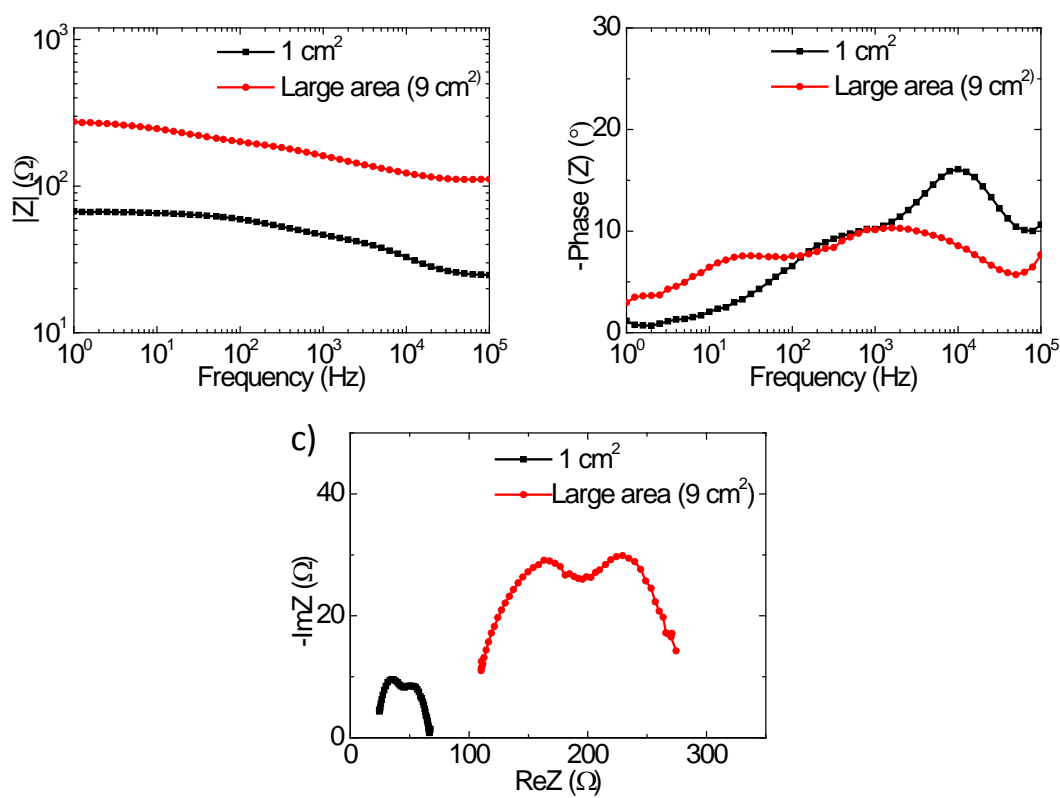

**Figure S20.** Electrochemical impedance spectroscopy spectra of the 1 cm$^2$-area (black dotted lines) (FTO/GO/rr-P3HT:PCBM/TiO$_2$/Pt/C-Nafion) and 9 cm$^2$-area photocathodes (red dotted lines) (ITO/GO/rr-P3HT:PCBM/TiO$_2$/Pt/C-Nafion) at 0 V *vs.* RHE and under 1.5AM illumination. Bode plots of the a) |Z| and b) of the -phase(Z) and c) the corresponding Nyquist plots.